\documentclass[aps,prb,twocolumn,groupedaddress]{revtex4}

\usepackage{overpic}

\usepackage{amsmath}
\usepackage{bm}

\usepackage{mathrsfs}

\usepackage{overpic}
\usepackage{mathrsfs}

\usepackage{subfig}
\usepackage{multirow}

\usepackage{afterpage}
\usepackage{float} 

\usepackage{mathtools}

\usepackage{afterpage}
\usepackage{float} 

\DeclareMathAlphabet{\mathscrbf}{OMS}{mdugm}{b}{n}

\begin{document} 

\title{Electronic structure and thermoelectric properties of n- and p-type SnSe \\ 
from first principles calculations}

\author{K. Kutorasinski}
\email[]{kamil.kutorasinski@fis.agh.edu.pl}
\author{B. Wiendlocha}
\author{S. Kaprzyk}
\author{J. Tobola}
\affiliation{AGH University of Science and Technology, Faculty of Physics and Applied Computer Science, Al. Mickiewicza 30, 30-059 Krakow, Poland}

\date{\today}

\begin{abstract}
We present results of the electronic band structure, Fermi surface and electron transport properties calculations in the orthorhombic $n$- and $p$-type SnSe, applying the Korringa-Kohn-Rostoker method and the Boltzmann transport approach. The analysis accounted for the temperature effect on crystallographic parameters in $Pnma$ structure as well as the phase transition to $CmCm$ structure at $T_c\sim 807~$K. Remarkable modifications of the conduction and valence bands were notified upon varying crystallographic parameters within the structure before $T_c$, while the phase transition mostly leads to the jump in the band gap value. The diagonal components of the kinetic parameter tensors (velocity, effective mass) and resulting transport quantity tensors (electrical conductivity $\sigma$, thermopower $S$ and power factor PF) were computed in a wide range of temperature ($15-900~$K), and hole ($p-$type) and electron ($n-$type) concentrations ($10^{17}-10^{21}$ cm$^{-3}$). 
SnSe is shown to have a strong anisotropy of the electron transport properties for both types of charge conductivity, as expected for the layered structure, with the generally heavier $p$-type effective masses, comparing to $n$-type ones. 
Interestingly, $p$-type SnSe has strongly non-parabolic dispersion relations, with the 'pudding-mold'-like shape of the highest valence band. 
The analysis of $\sigma$, $S$ and PF tensors indicates, that the inter-layer electron transport is beneficial for thermoelectric performance in $n$-type SnSe, while this direction is blocked in $p$-type SnSe, where in-plane transport is preferred. 
Our results predict, that the $n$-type SnSe is potentially even better thermoelectric material than the $p$-type one. 
Theoretical results are compared with the single crystal $p$-SnSe measurements, and a good agreement is found below $600~$K. 
The discrepancy between the computational and experimental data, appearing at higher temperatures, can be explained assuming an increase of the hole concentration vs. $T$, which is correlated with the experimental Hall data. 
\end{abstract}

\maketitle

\section{Introduction}
\label{sec:Introduction}
Thermoelectric (TE) conversion in crystalline solids constantly attracts the interest of scientists, not only due to the increasing performance of energy harvesting systems, but also due to the fact that the conventional and 'well-known' thermoelectric materials, such as Pb$X$ ($X=$S, Se, Te)~\cite{PbTe-Heremans,PbSeS-Snyder,PbSeS-Story} or 
Bi$_2$Te$_3$ (e.g.\cite{Bi2Te3-Cava}), still surprise by their novel and remarkable physical behaviors. 
It is commonly accepted, that TE figure of merit $zT=\sigma S^2/\kappa T$ well captures basic transport properties of the material at given temperature $T$, i.e. electrical conductivity ($\sigma$), Seebeck coefficient ($S$) and thermal conductivity ($\kappa$), and is conveniently expressed in dimensionless units. As all these transport quantities apparently depend on temperature and carrier concentration ($n$ or $p$), the maximum of $zT$ is expected when properly correlating both intrinsic electron transport properties of the system (e.g. the band gap magnitude) with temperature range, and doping level with the hole or electron concentration, to achieve $p-$ or $n-$type materials, respectively. But, getting a better insight into atomic-level connections among crystal stability, electronic and lattice properties of TE systems generally allow for a more convinced interpretation of the complex transport phenomena. 

Recent experimental work~\cite{SnSeNature} reported $zT\sim2.6$ at $T\sim 920$~K, along one of the axis in the single crystal $p-$ type SnSe, which classified this well-known semiconductor as a very promising TE material.
However, further results~\cite{snse-Snyder,snse-Lenoir} revealed much lower $zT$ for polycrystalline samples, showing that SnSe is a rather complex system.

\begin{figure}[b]
\centering
\includegraphics[height=0.135\textheight]{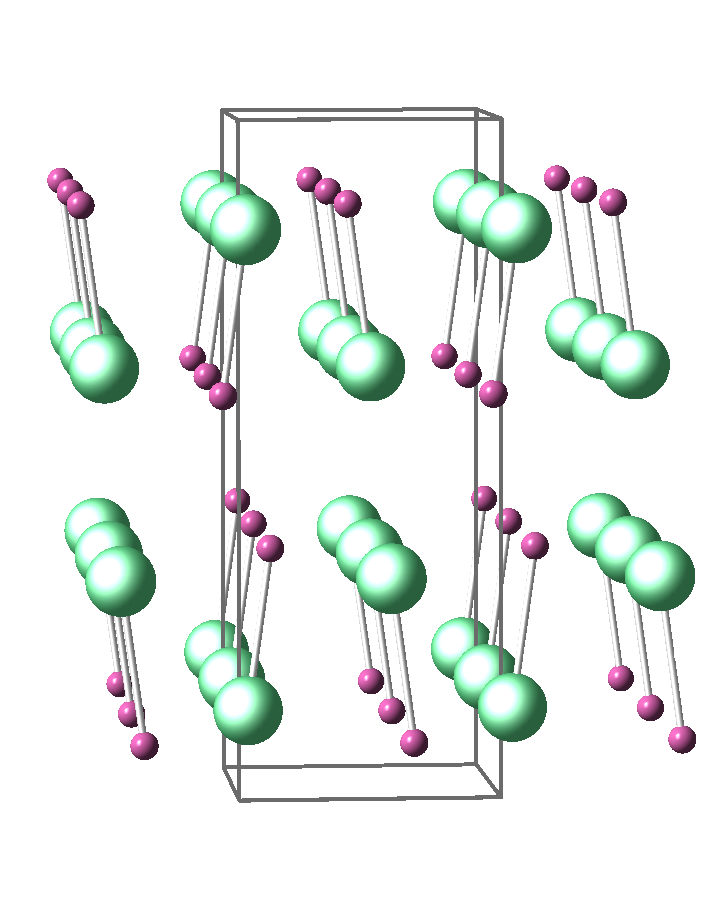}
\includegraphics[height=0.135\textheight]{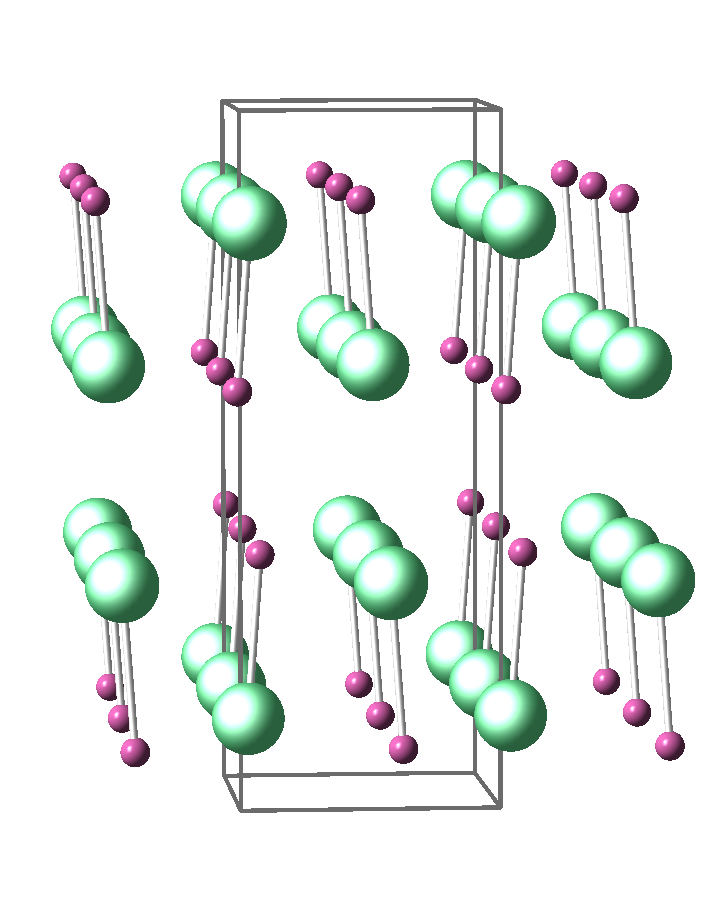}
\includegraphics[height=0.135\textheight]{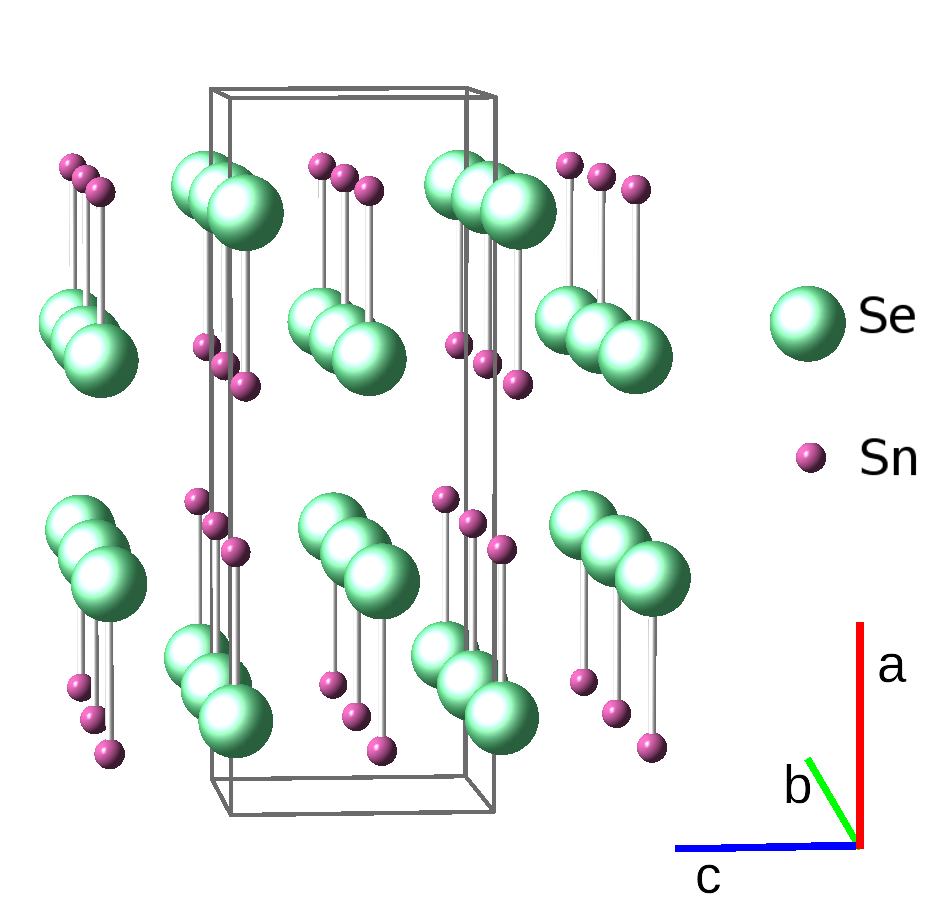}
\caption{(Color online) Crystal structure of SnSe (Se - large, green balls, Sn- small, magenta). Left: LT phase ({\it Pnma} at 295 K); center: MT ''phase'' ({\it Pnma} at 790 K); right: HT phase ({\it Pbmm} above 807 K). }
\label{fig:1}
\end{figure}

\begin{table}[b]
\caption{Crystallographic data~\cite{vonSchnering} of the SnSe compound, measured at different temperatures $T_{\rm expt}$, 
used in our calculations for selected temperature ranges $T_{\rm calc}$.}
\centering
\begin{tabular}{lccrll}
\hline\hline
Phase, & Space  & Lattice   & \multicolumn{3}{c}{ \ \ Atomic positions} \\
$T$ range & group & const.(\AA) & &Sn&Se \\
\hline
\multirow{3}{*}{\begin{tabular}[c]{@{}l@{}}Low-T. (LT) \\ $T_{\rm expt}=295$ K  \\ $T_{\rm calc} < 550$~K \end{tabular}}  & {$Pnma$}  &  $a=11.501$ &x:& 0.6186& 0.3551  \\
 &  No. 62 &  $b=4.153$ &\ \ \ \ y:& 0.25& 0.25  \\
 &   &  $c=4.445$ &z:& 0.1032& 0.4818  \\[0.8ex]
\hline
 \multirow{3}{*}{\begin{tabular}[c]{@{}l@{}}Mid-T. (MT) \\ $T_{\rm expt}=790$ K \\ $T_{\rm calc}$=550-807~K \end{tabular}}  & {$Pnma$}  &  $a=11.620$ &x:& 0.6230& 0.3553  \\
  & No. 62  &  $b=4.282$ &y:& 0.25& 0.25 \\
  &   &  $c=4.334$ &z:& 0.0584& 0.4878 \\[0.8ex]
\hline
\multirow{3}{*}{\begin{tabular}[c]{@{}l@{}}High-T. (HT) \\ $T_{\rm expt}=829$ K \\ $T_{\rm calc} > 807$~K \end{tabular}}   & $Pbmm$\footnotemark[1]   & $a=11.62$ &x:& 0.6248 & 0.3558 \\
 &  No. 51\footnotemark[1]  &  $b=4.282$ &y:& 0.25& 0.25 \\
 &   &  $c=4.293$ &z:& 0.00& 0.50\\[0.8ex]
\hline\hline
\end{tabular}
\footnotetext[1]{Transformed from $CmCm$ (No. 61)}
\label{table:structure}
\end{table}

SnSe is a layered material, with an orthorhombic symmetry of the unit cell, and its crystal structure strongly depends on temperature. First of all, SnSe exhibits a crystallographic phase transition at $T_c = 807$~K.\cite{vonSchnering} Below $T_c$, its unit cell can be described in the {\it Pnma} simple orthorhombic space group No. 62, with 4 chemical formulas (8 atoms) in the unit cell (see, Fig.~\ref{fig:1}). Crystal structure parameters of this 'low temperature' (i.e. below $T_c$) phase depend on temperature as well~\cite{vonSchnering}. Since the temperature dependence of the transport properties (e.g. thermopower) is discussed in our work, those changes in crystal structure were also taken into account. Accordingly, three cases are considered: low-temperature (LT) phase, described by the experimental crystallographic data taken at $T_{\rm expt} = 295$~K, and used in our calculations to represent SnSe in the temperature range $T_{\rm calc} = 10 - 550$~K; mid-temperature (MT) phase ($T_{\rm expt} = 790$~K 
and $T_{\rm calc} = 550-807$~K); and high-temperature (HT) phase ($T_{\rm expt}=829$ K and $T_{\rm calc} > 807$~K), as displayed in the Table~\ref{table:structure}. 
Above $T_c$, the high-temperature phase has an orthorhombic $c$-base centered {\it CmCm} space group (No. 63). The transition from the simple to the centered unit cell reduces the number of atoms in the primitive cell, from 8 to 4, changing the shape and size of the Brillouin zone (BZ). Thus, it becomes impossible to directly compare the electronic dispersion relations between the LT/MT and HT phases, if using the HT centered unit cell. Therefore, to allow for a better understanding, and easier analysis of the role of the phase transition in the 
evolution of the electronic structure and transport properties of SnSe, we transformed the HT SnSe centered unit cell into the equivalent {\it Pbmm} simple orthorhombic one (No. 51). {\it Pbmm} unit cell has the same number of atoms and similar alignment of the unit cell axes, as the structure before the transition (i.e. $a$ axis is the longest, $b$ is the shortest, and $c$ is in between, see, Fig.~\ref{fig:1}). 
With this transformation, the Brillouin zone and location of the high symmetry points remain similar in both phases 
(small changes are only in BZ dimensions, due to lattice parameters' variation). 
In our computations, the actual values of lattice parameters for HT phase were taken from the high temperature 
($T = 829$~K) neutron measurements~\cite{vonSchnering}, and are shown in the Table.~\ref{table:structure}. 
It is worth noting that the crystal structure of SnSe evolves smoothly with temperature (even while crossing $T_c$) and the unit cell dimensions, as well as interatomic distances, change rather continuously with temperature (see, Table~\ref{table:structure} and Ref.~\onlinecite{vonSchnering}). The crystal structure parameter, which exhibits the most rapid change at $T_c$, is the $z$ parameter positioning Sn and Se atoms.

Recently, first principles calculations of transport properties of $p$-type SnSe were reported in Ref.~\onlinecite{snse-arxiv}. In our work we give more extended analysis of the electronic structure, effective masses and transport properties of both, $n$- and $p$-type, SnSe. The constant relaxation time approximation is used, and phonon drag effects are not discussed.

The paper is organized as follows. In Sec.~\ref{sec2} theoretical and computational details are presented. Sec.~\ref{sec3a} describes in details the electronic band structure and the Fermi surface of the valence and conduction states, together with the effective mass and the transport function analysis. Sec.\ref{sec:thermo} discusses anisotropic (single crystal) and isotropic (polycrystalline-like) transport properties in function of temperature and carrier concentration, and comparison with experimental data is done. Sec.~\ref{sec:Conclusions} contains summary and conclusions of the work. In Appendix~\ref{sec:append1}, formulas for averaging of the anisotropic transport coefficients, in order to compute their isotropic (polycristalline) analogs, are derived. Appendix~\ref{sec:append2} contains complementary data.

\section{Theoretical Details}\label{sec2}

\subsection{Thermopower}

 The Boltzmann transport theory\cite{Ashcroft,Thonhauser,Madsen}, successfully applied to study transport properties of various crystalline materials\cite{Chaput,kk13,KKBW14} is used to calculate the energy-dependent electrical conductivity $\bm\sigma(\mathscr{E})$ (so-called transport function, TF): 
\begin{equation}
\bm\sigma(\mathscr{E})=e^2\sum_n\int\frac{d \mathbf{k}}{4 \pi^3}\tau_n(\mathbf{k})\mathbf{v}_n(\mathbf{k})\otimes\mathbf{v}_n(\mathbf{k})\delta(\mathscr{E}-\mathscr{E}_n(\mathbf{k})).
\end{equation}
Symbol $\otimes$ represents outer product (Kronecker product) of two vectors, $\mathbf{v}_n(\mathbf{k})=\nabla_k\mathscr{E}_n(\mathbf{k})$ and $\tau_n(\mathbf{k})$ is a velocity and life time of electrons on band $\mathscr{E}_n(\mathbf{k})$,  respectively. 
The transport function tensor has to be reformulated into the form, that is more convenient for the numerical computation. With the use of the constant relaxation time approximation ($\tau_n(\mathbf{k})=\tau_0$), and after changing 3D $\mathbf{k}$-space integration to the 2D surface integration ($\mathscr{E}_n(\mathbf{k})\rightarrow S_n(\mathscr{E})$) it takes the form:
\begin{equation}\label{eq:sig}
\bm\sigma_{\tau}(\mathscr{E})=\tau_0\frac{e^2}{\hbar}\sum_n\int\limits_{S_n(\mathscr{E})}\frac{dS}{4\pi^3}\frac{\mathbf{v}(S_n(\mathscr{E}))\otimes\mathbf{v}(S_n(\mathscr{E}))}{|\mathbf{v}(S_n(\mathscr{E}))|}.
\end{equation}
TF is directly related to the macroscopic transport coefficients, like thermopower, electronic part of thermal conductivity or electrical conductivity. Within this approach\cite{Ashcroft} the two basic transport tensors (electrical conductivity $\bm \sigma_e$ and thermopower $\bm S$) can be expressed as
\begin{equation}\label{eq:s}
\bm \sigma_e=\mathscrbf{L}^{(0)},\ \ \  \bm S=-\dfrac{1}{eT}\dfrac{\mathscrbf{L}^{(1)}}{\mathscrbf{L}^{(0)}}, 
\end{equation}
where
\begin{equation}\label{eq:L}
\mathscrbf{L^{(\alpha)}}=\int{d\mathscr{E}\left(-\dfrac{\partial f}{\partial \mathscr{E}}\right)(\mathscr{E}-\mu_c)^\alpha}\bm\sigma(\mathscr{E}).
\end{equation}
The value of the chemical potential $\mu_c=\mu_c(T,n_d)$ depends on the temperature $(T)$, carrier concentration and the type of conductivity ($n$, $p$) (see, Sec.~\ref{potchem}).

In case of the anisotropic structure, like the orthorhombic SnSe, transport function tensor (and thermopower as well) has three linearly independent elements, and in this work, diagonal elements (i.e. $S_{xx}$ , $S_{yy}$, $S_{zz}$) are calculated. For the polycrystalline sample without texture, those three elements have to be averaged in a proper way, to obtain the macroscopically isotropic thermopower and power factor (PF). The relevant formulas are given here and derived in the Appendix~\ref{sec:append1}:
\begin{equation}
\label{eq:Save}
S_{\rm avg}=\frac{S_{xx}\sigma_{xx}+S_{yy}\sigma_{yy}+S_{zz}\sigma_{zz}}{\sigma_{xx}+\sigma_{yy}+\sigma_{zz}},
\end{equation}
\begin{equation}
\label{eq:PFave}
{\rm PF}_{\rm avg}=\frac{1}{3}(S_{xx}^2\sigma_{xx}+S_{yy}^2\sigma_{yy}+S_{zz}^2\sigma_{zz}).
\end{equation}

Since the power factor, in the constant relaxation time approximation, depends linearly on the relaxation time $\tau$ (not calculated here), consequently PF divided by $\tau$ is presented.

\subsection{DOS and effective mass}

In a similar way to the transport function, the density of states (DOS) is calculated:
\begin{equation}
g(\mathscr{E})=\sum_n\int\limits_{S_n(\mathscr{E})}\frac{dS}{4\pi^3}\frac{1}{|\nabla_k\mathscr{E}_n(\mathbf{k})|}.
\end{equation}
With this definition $g(\mathscr{E})$ has the units of eV$^{-1}$m$^{-3}$, i.e. it includes the volume of the unit cell.

The DOS function is closely connected to the DOS effective mass, $m^*$. Here, energy dependent $m^*$ is calculated using the formula\cite{KKBW14} 
\begin{equation}
\label{eq:mass}
m_{\rm DOS}(\mathscr{E})=m_em^*_{\rm DOS}(\mathscr{E})=\hbar^2\sqrt[3]{\pi^4g(\mathscr{E})g'(\mathscr{E})}.
\end{equation}

Alternatively, effective mass can be computed by integrating the effective mass tensor, over the isoenergetic surfaces:

\begin{equation}
\label{eq:mass2}
m_{ij}(\mathscr{E})=m_em^*_{ij}(\mathscr{E})=\left.\displaystyle\int\limits_{S_n(\mathscr{E})}\left[\bm{M}_{ij}\right]dS \right/\displaystyle\int\limits_{S_n(\mathscr{E})}dS,
\end{equation}
where the effective mass tensor is defined as\cite{Ashcroft}
\begin{equation}
\label{eq:masstensor}
\left[\bm{M}_{ij}\right]^{-1}=\frac{1}{\hbar^2}\frac{\partial^2\mathscr{E}}{\partial k_i\partial k_j}
\end{equation}

Using Eq.~\ref{eq:mass2} and \ref{eq:masstensor} one may discuss the direction dependence of the effective mass. 
In case of an orthorhombic structure with orthogonal axes, $\left[\bm{M}_{ij}\right]$ tensor is diagonal, and components $m^*_{xx}$, $m^*_{yy}$, $m^*_{zz}$ are computed. 

The isotropic band effective mass is determined by the geometrical mean 
\begin{equation}
\label{eq:massall}
m^*_{iso}=\sqrt[3]{m^*_{xx}m^*_{yy}m^*_{zz}}.
\end{equation}
Both ways of effective mass calculations (DOS effective mass from Eq.~\ref{eq:mass} and band effective mass from Eq.~\ref{eq:massall}) give the same results only for parabolic bands. Difference of those two results gives opportunity to estimate the importance of non-parabolicity of the electronic band structure.

The aforementioned effective masses correspond to the $T = 0$~K temperature. The temperature effects can be taken into account via the Fermi-Dirac distribution function in a similar way, as in the thermopower calculation in Eq.~\ref{eq:s}. The actual number of 'active' (conducting) electrons, in temperature $T$, can be represented as $n_{active}=\int d\mathscr{E} g(\mathscr{E})\left(-\frac{\partial f}{\partial \mathscr{E}}\right)$ and the effective mass of active electrons, as a function of temperature, can be determined from
\begin{equation}
\label{eq:mass_n_T}
m^*(T,n_d)=\frac{\displaystyle\int d\mathscr{E} m^*(\mathscr{E})g(\mathscr{E})\left(-\frac{\partial f}{\partial \mathscr{E}}\right)}{\displaystyle\int d\mathscr{E} g(\mathscr{E})\left(-\frac{\partial f}{\partial \mathscr{E}}\right)}.
\end{equation}
where $n_d$ is the carrier concentration, at which $m^*$ is calculated (see, Sec. \ref{potchem}).
Note, that this analysis requires an assumption that effective mass is well defined, which for strongly non-parabolic bands may not be valid. 

\subsection{Doping and chemical potential\label{potchem}}
SnSe is an intrinsic semiconductor, where the Fermi energy lies inside the gap. As already mentioned, transport properties of the intrinsically (defect-doped) $p$-type single crystal~\cite{SnSeNature}, polycrystal~\cite{snse-Lenoir}, and $p$-type Ag-doped polycrystal~\cite{snse-Snyder} samples were reported. To simulate the behavior of the system after doping, we use the rigid band model~\cite{RBM}. In this approach, additional number $n_d$ of electrons or holes is the control parameter, added to mimic $n-$type (positive $n_d$) or $p-$type (negative $n_d$) behavior, and chemical potential $\mu_c=\mu_c(T,n_d)$ needed in Eq.~(\ref{eq:L}) is calculated using the formula
\begin{equation}
n+n_d=\int d\mathscr{E}g(\mathscr{E})\dfrac{1}{1+\exp\left(\frac{\mathscr{E}-\mu_c(T,n_d)}{k_BT}\right)}.
\label{eq:potchem}
\end{equation}
 $n$ is the total number of valence electrons in the system (which is 10 per f.u. in case of SnSe) and the integral is taken from the bottom of the valence bands.

\subsection{Band structure computational details}

Electronic band structure calculations were performed using the full potential Korringa-Kohn-Rostoker\cite{Kohn54,Butler76,Kaprzyk82,Kaprzyk90,Bansil99,Stopa04} (KKR) method, within the scalar relativistic approach~\cite{KOELLING,Ebert}. 
Local density approximation (LDA) parametrization of Perdew and Wang~\cite{PW92} was employed. 
The self-consistent cycle was repeated until the difference between the input and output potentials was less than $1$~mRy in any point inside the unit cell. 
Isoenergetic surfaces $S_n(\mathscr{E})$ were obtained with the use of marching cube algorithm~\cite{Lorensen} on a mesh consisting of $80\times80\times80$ voxels. To improve visualization, vertex normal~\cite{Gouraud} and vertex color techniques were also used. 
All energy dependent functions were calculated with a resolution of $2.5$~meV. The transport function was additionally interpolated between energy mesh points, using spline functions, which allows to obtain converged results for temperatures above 10 K, in the concentration range of $10^{17}-10^{21}$~cm$^{-3}$. 

\section{ Results }
\label{sec:results}

\subsection{Electronic Structure}\label{sec3a}

\begin{figure*}[t!]
\centering
\includegraphics[height=0.149\textheight]{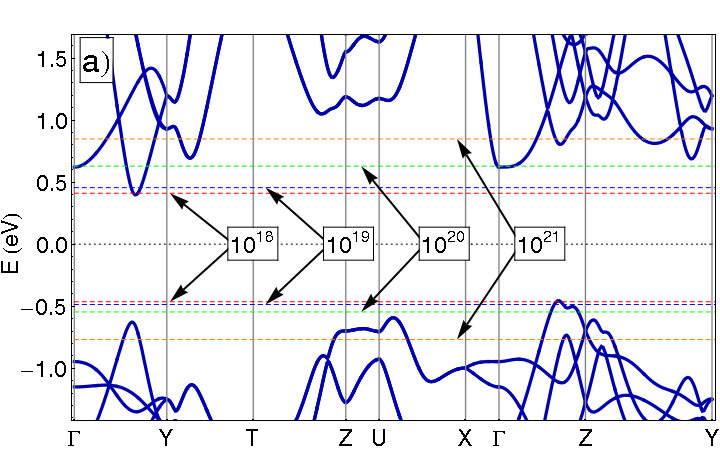}
\includegraphics[height=0.149\textheight]{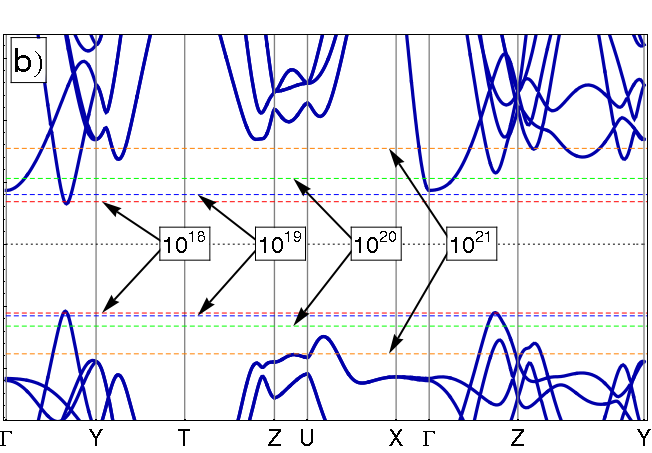}
\includegraphics[height=0.149\textheight]{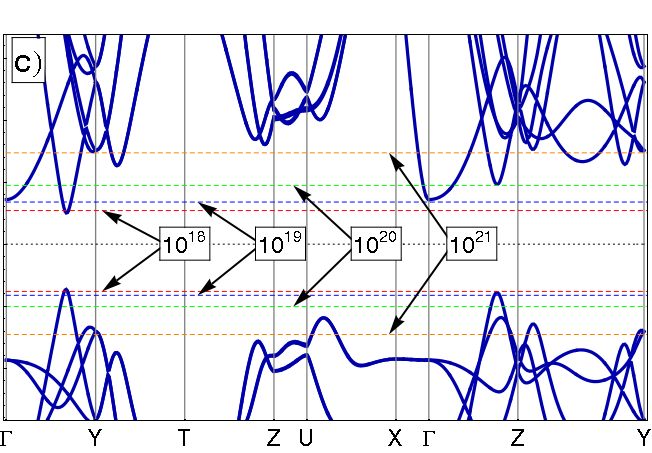}
\includegraphics[height=0.123\textheight]{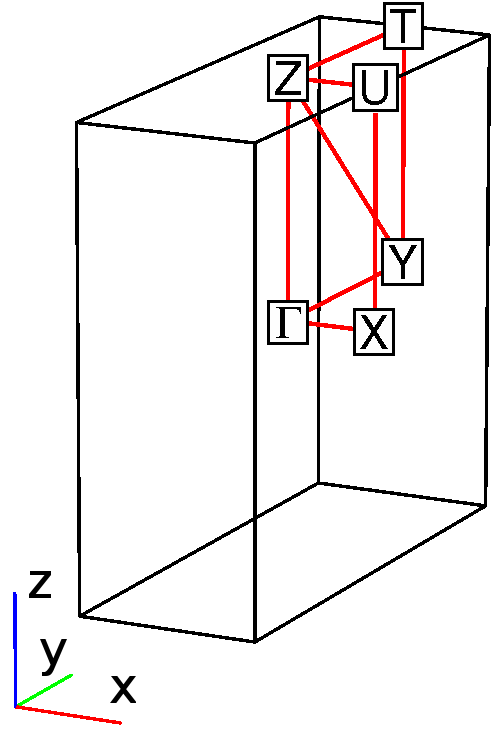}
\caption{(Color online) Electronic band structure of the LT (a), MT (b), and HT (c) phases of SnSe, computed along the high symmetry directions, as shown in the orthorhombic Brillouin zone. Horizontal lines mark the Fermi energy positions ($\mu_c$ at T = 0 K)} for the electron/hole concentration
(in cm$^{-3}$) of $n=10^{18}$ (red), $n=10^{19}$ (blue), $n=10^{20}$ (green) and $n=10^{21}$ (orange), respectively.
\label{fig:bands1}
\end{figure*}

Electronic dispersion relations are shown in Fig.~\ref{fig:bands1}, for the LT (Fig.~\ref{fig:bands1}a), MT (Fig.~\ref{fig:bands1}b) and HT (Fig.~\ref{fig:bands1}c) phases\footnote{We use the term {\it phase} to name the three different crystal unit cells of SnSe, even if, in fact, LT and MT have the same space group, so are not exactly different crystalline phases.} of SnSe. As expected, local density approximation (LDA), used in this work, underestimated the band gaps. Since the gap value is the important parameter in transport properties calculations, especially in elevated temperatures, the computed band gaps were expanded to mimic the experimental ones (see Tab.~\ref{table:tabgap}). 
For the LT phase, the band gap value was set to $E_g^{\rm exp} =0.86$~eV~\cite{SnSeNature}. In MT and HT phases, where experimental data are not available, the calculated gap values were extrapolated proportionally as in the LT phase, i.e. we assume that the LDA underestimation is proportional to the real gap value. 
It is worth noting, that the calculated values of the energy gaps for the LT and MT phases are almost the same, whereas the HT phase has much smaller gap (see, Tab.~\ref{table:tabgap}).
 To verify whether the reduction of the band gap is related to the change in atomic positions or the unit cell parameters, additional calculations were performed for the MT phase, using (i) the MT atomic parameters and HT lattice constants, and (ii) MT lattice constants and HT atomic parameters. The resulting LDA band gaps were: 0.465 eV and 0.355 eV, respectively, thus we found, that the gap value is controlled mainly by the Sn and Se atomic positions.
Recent calculations~\cite{snse-arxiv}, using the GW method, reported $E_g=0.829$~eV for the LT phase and 0.46~eV for HT, thus reduction of $E_g$ after the phase transition was also found.

\begin{table}[b]
\caption{Calculated (LDA) and experimental~\cite{SnSeNature} values of the energy band gap (in eV) in SnSe for  the three structures, LT, MT and HT.}
\centering
\begin{tabular}{lccc}
\hline\hline
Structure  & \ LDA \ & \ Experimental \ & \  Used \ \\
\hline
\ LT  \ & 0.474  & 0.86\footnotemark[1]  & 0.86  \\
\ MT  & 0.487  & no data  & 0.87  \\
\ HT  \ & 0.350  & no data   & 0.64    \\ [.3ex]
\hline\hline
\end{tabular}
\label{table:tabgap}
\end{table}

\begin{figure*}[hbt]
\centering
\begin{tabular}{lcr}
\subfloat[$n$-type LT]{%
\includegraphics[width=0.14\textwidth]{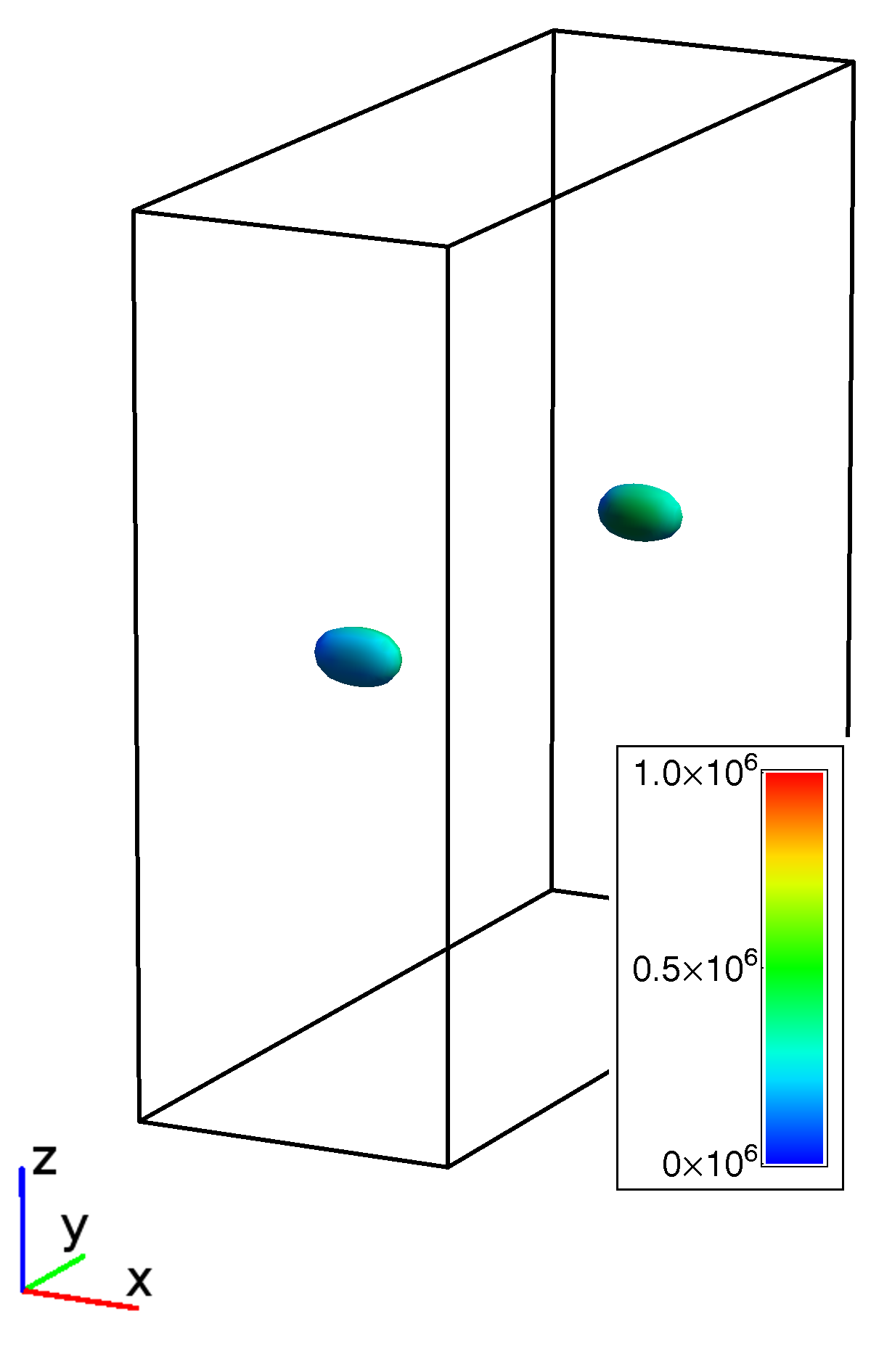}
\includegraphics[width=0.14\textwidth]{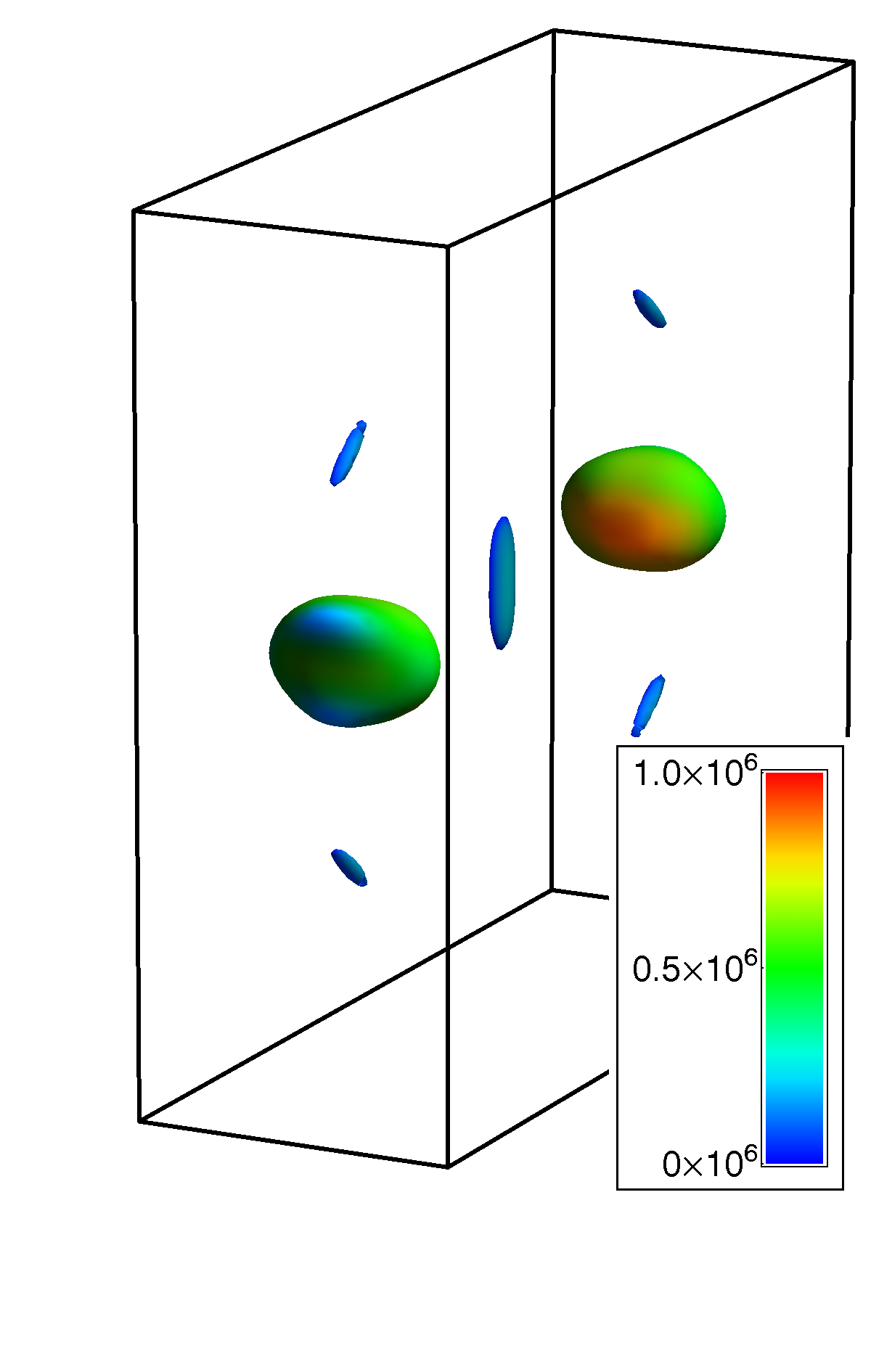}
}& \ \ \ \ \ \ \ \ 
\subfloat[$n$-type MT]{%
\includegraphics[width=0.14\textwidth]{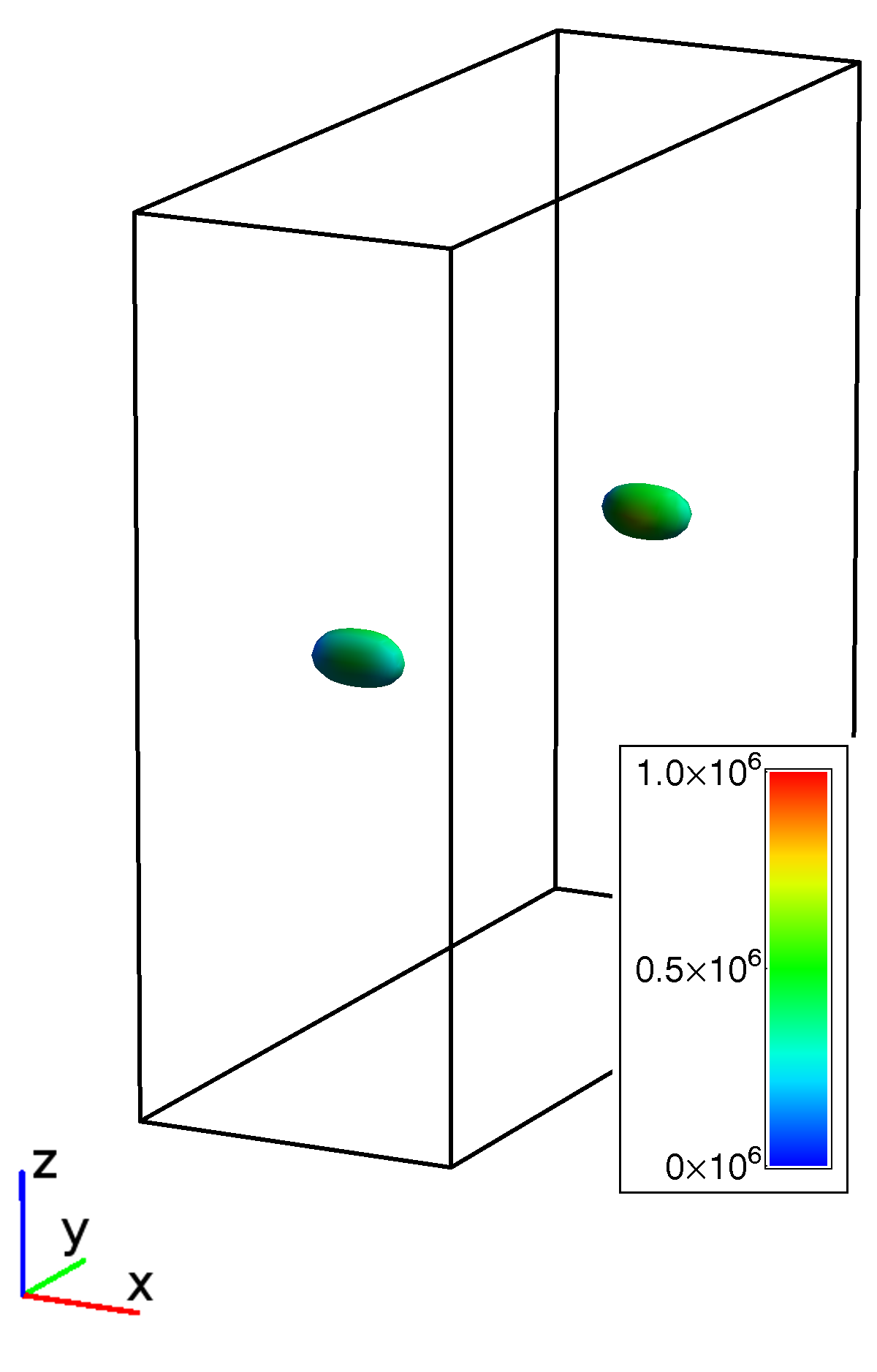}
\includegraphics[width=0.14\textwidth]{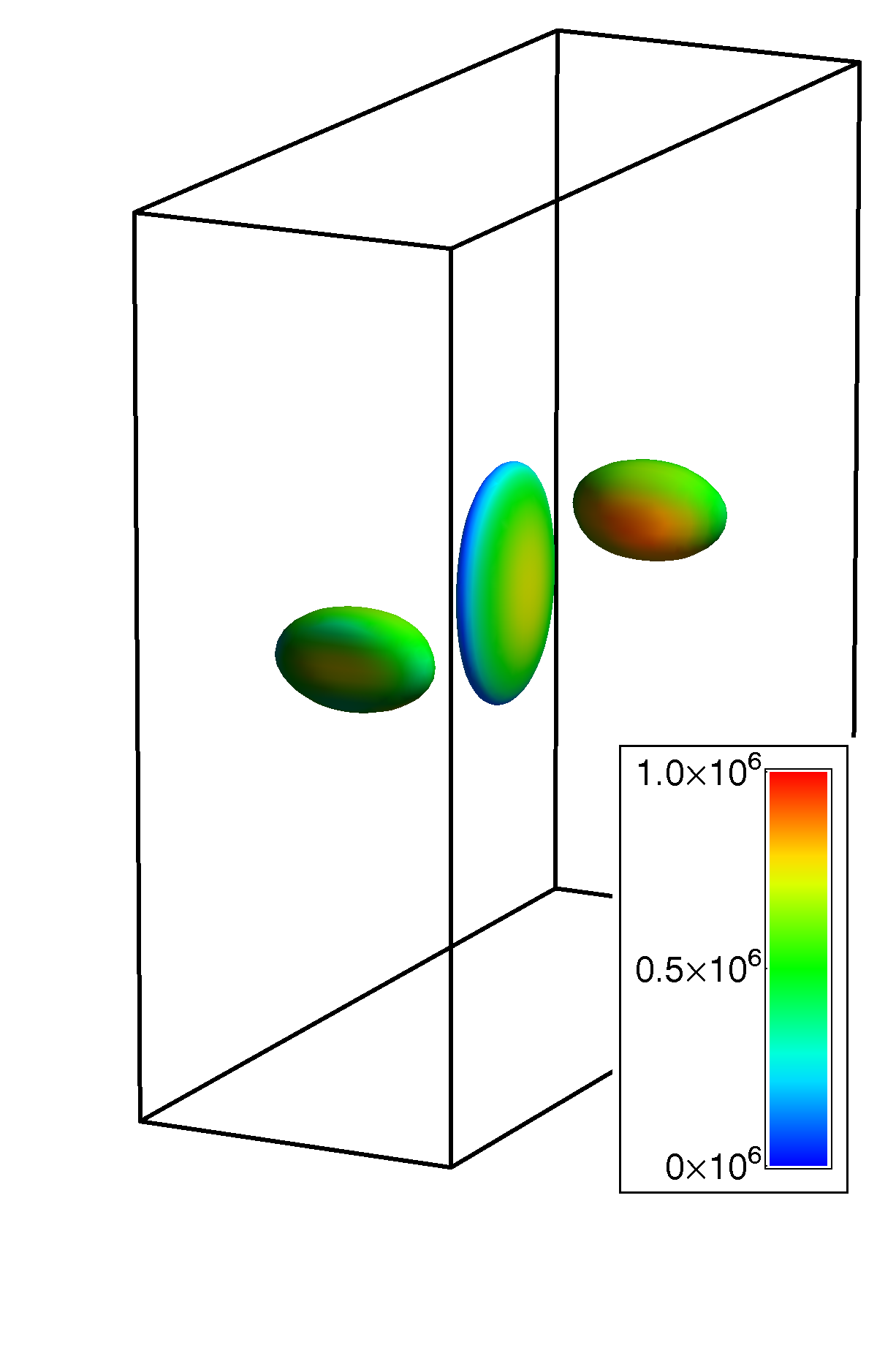}
}& \ \ \ \ \ \ \ \ 
\subfloat[$n$-type HT]{%
\includegraphics[width=0.14\textwidth]{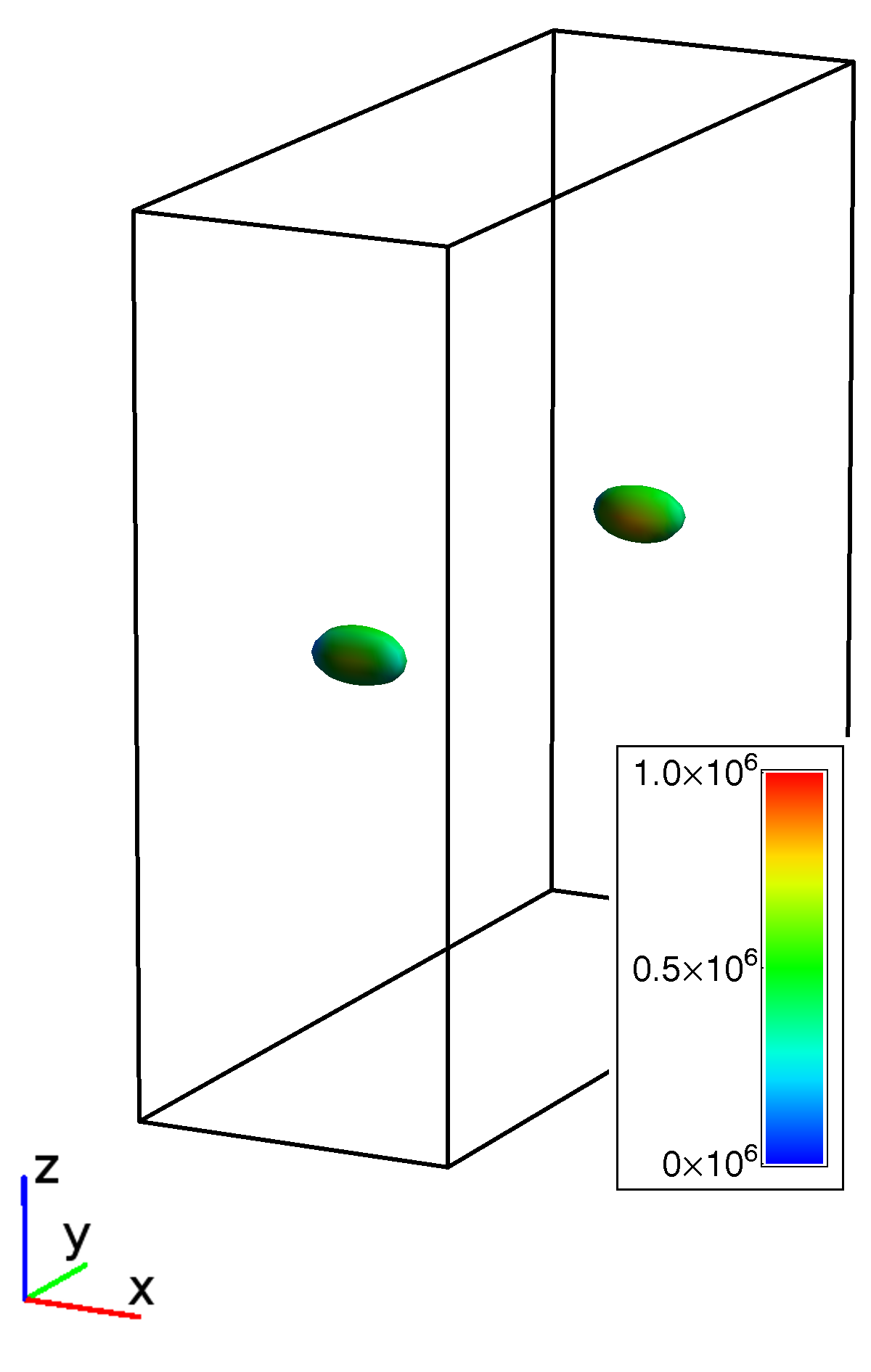}
\includegraphics[width=0.14\textwidth]{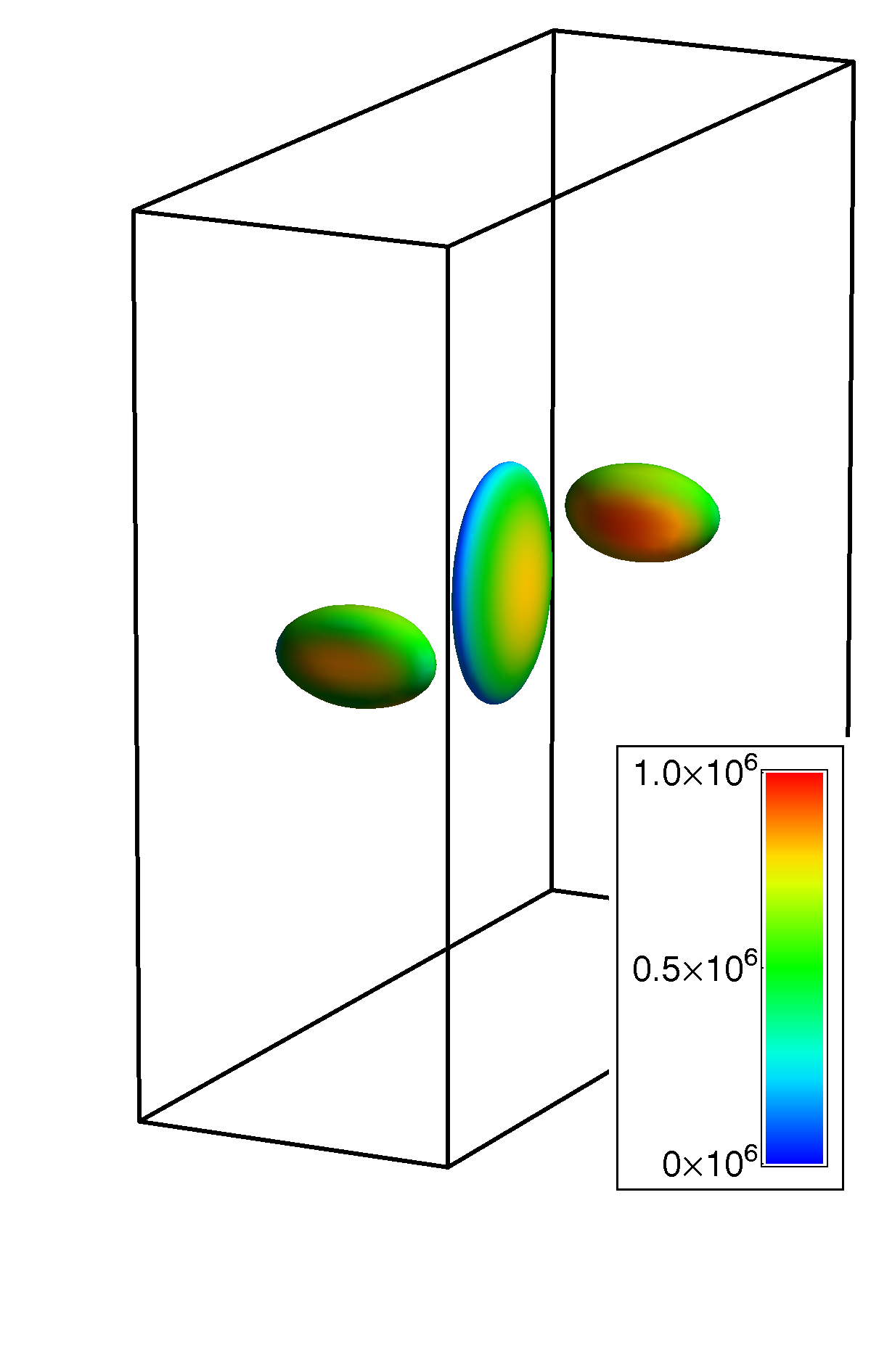}
 }\\
\subfloat[$p$-type LT]{%
\includegraphics[width=0.14\textwidth]{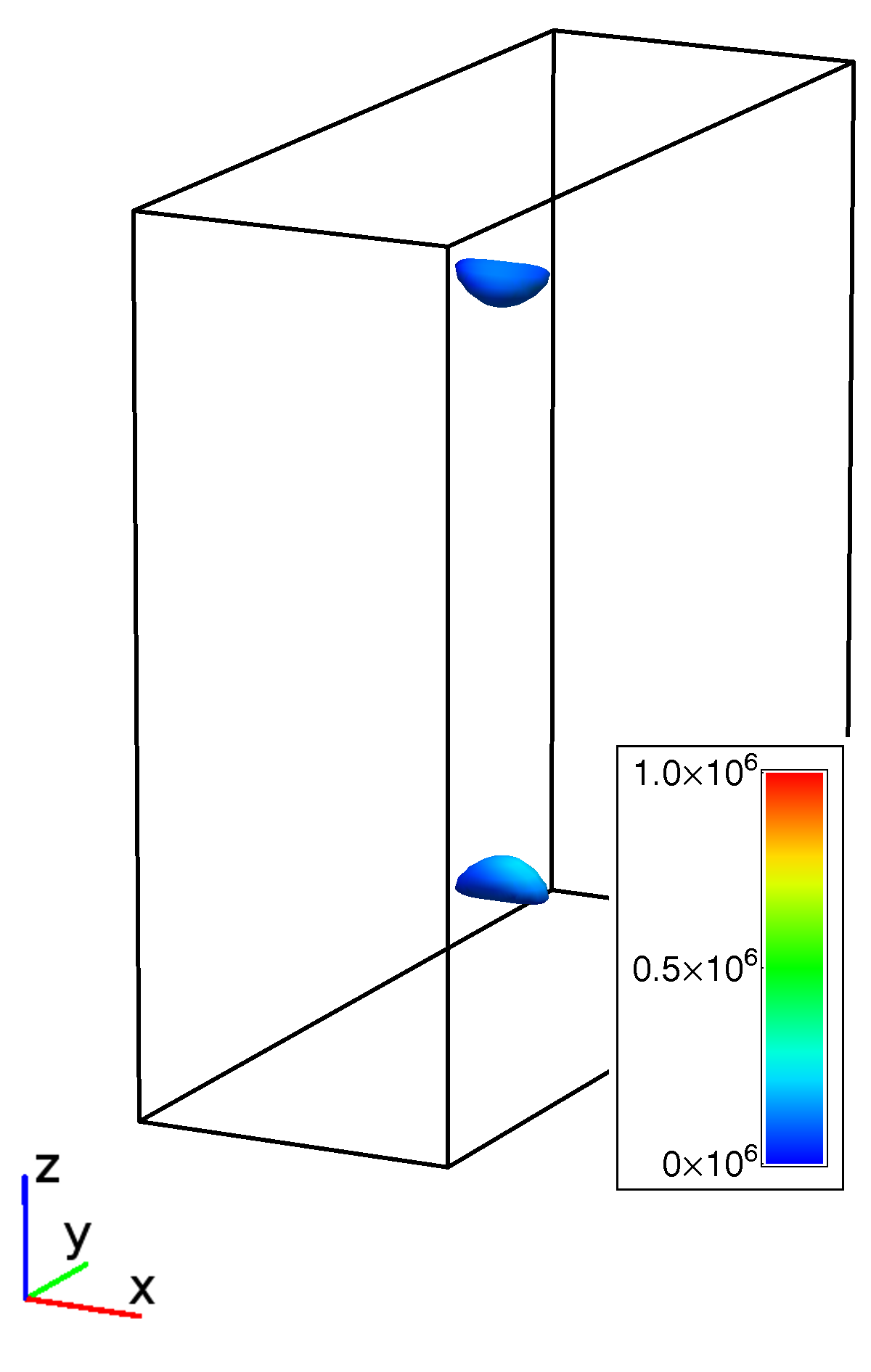}
\includegraphics[width=0.14\textwidth]{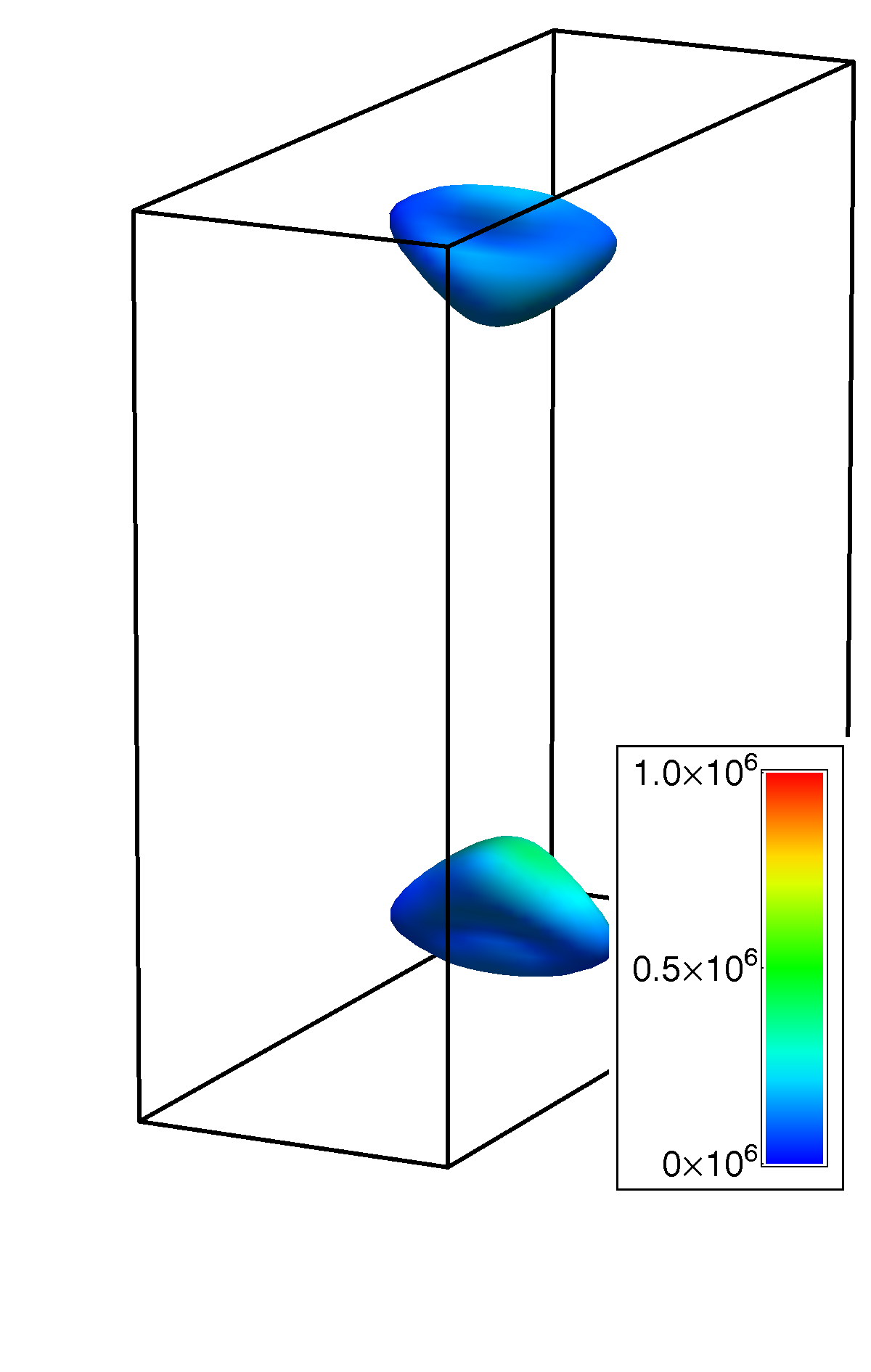}
}& \ \ \ \ \ \ \ \ 
\subfloat[$p$-type MT]{%
\includegraphics[width=0.14\textwidth]{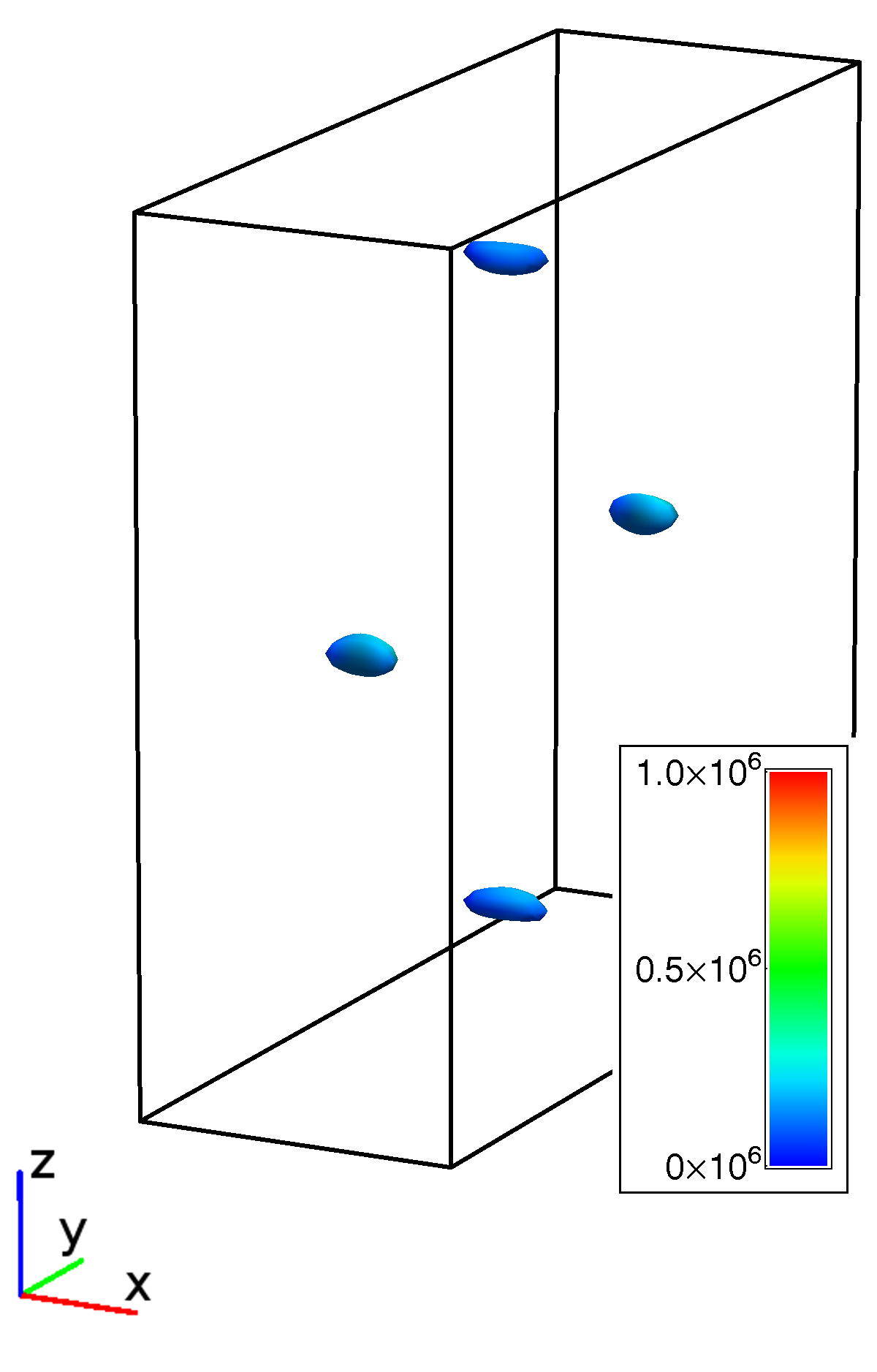}
\includegraphics[width=0.14\textwidth]{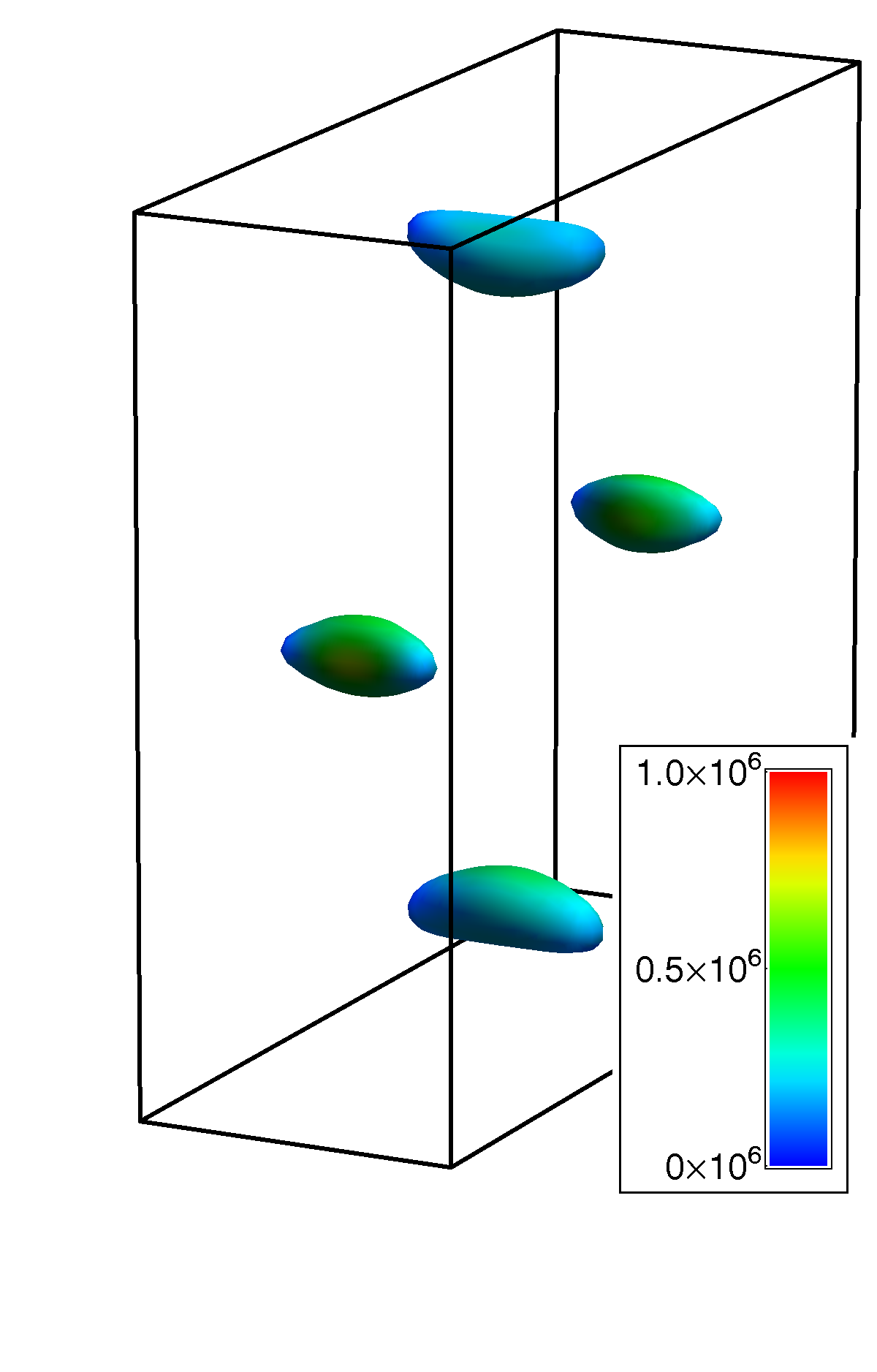}
}& \ \ \ \ \ \ \ \ 
\subfloat[$p$-type HT]{%
\includegraphics[width=0.14\textwidth]{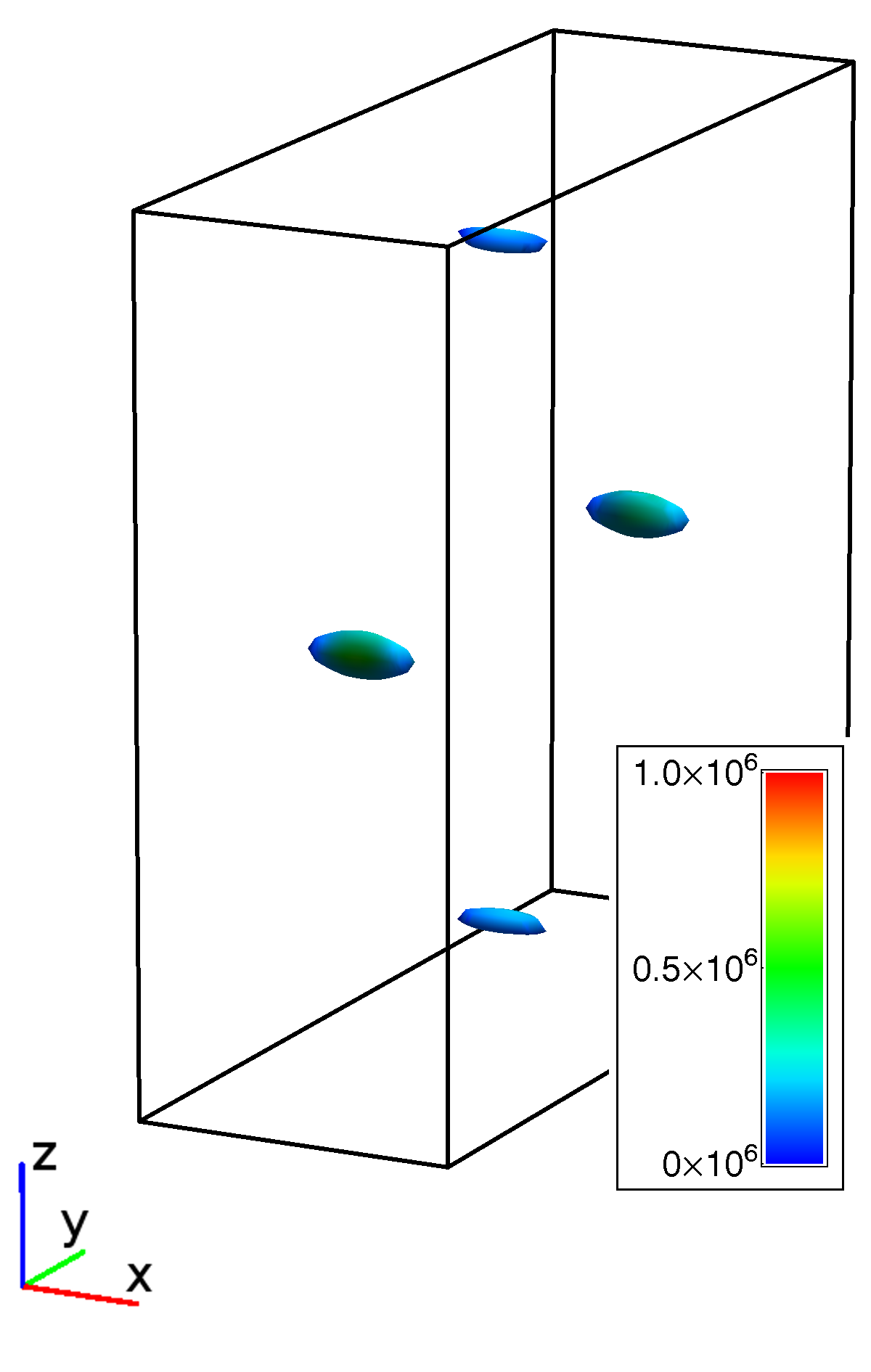}
\includegraphics[width=0.14\textwidth]{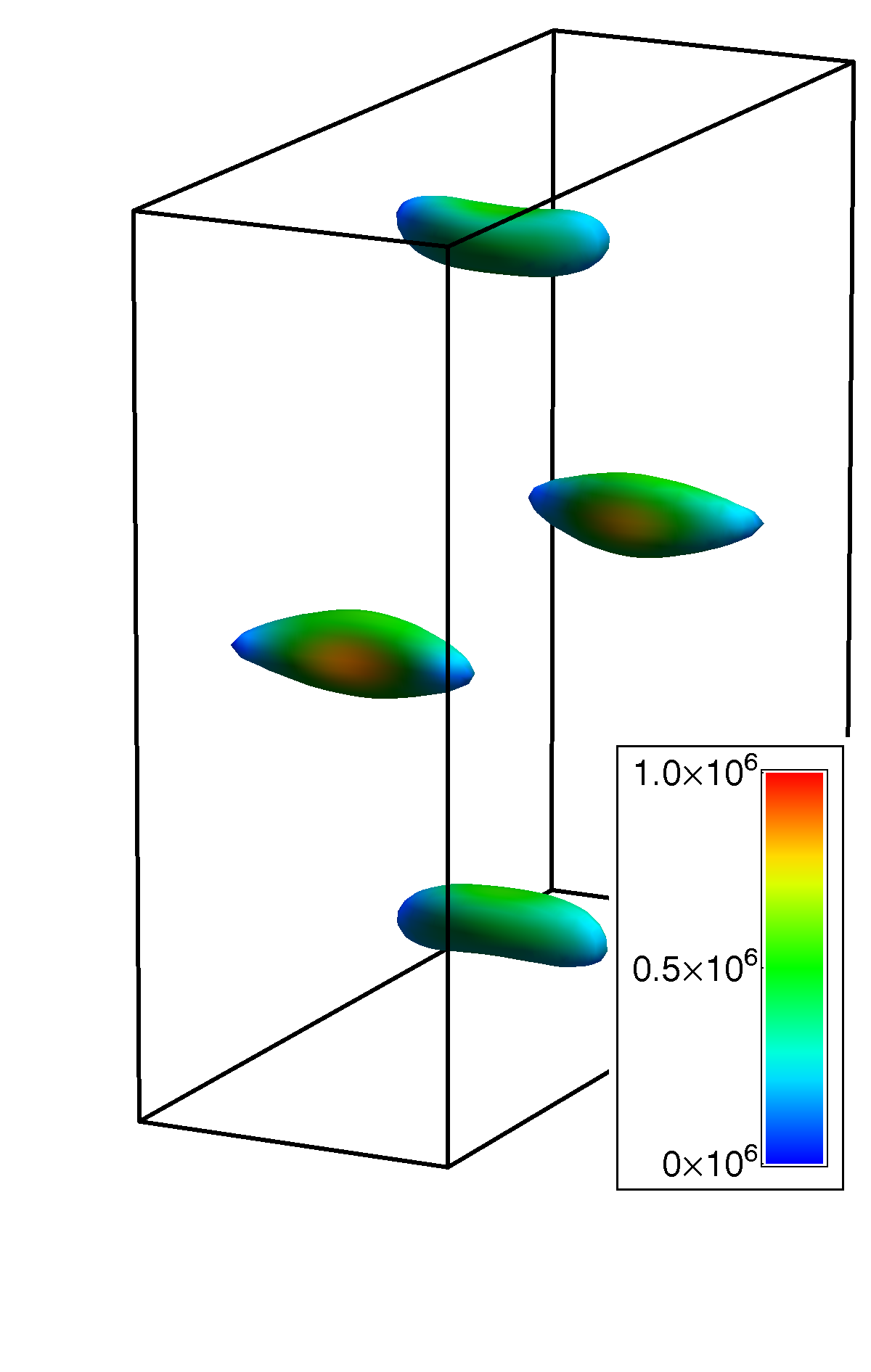}
 }
\end{tabular}
\caption{(Color online) Fermi surfaces of the LT (a,d), MT (b,e) and HT (c,f) phases of SnSe, for the $n-$type (a-c) and $p-$type (d-f) doping, for the carrier concentration (in cm$^{-3}$) $10^{19}$ (left sub-panel) and $10^{20}$ (right sub-panel). Electron velocities (in m/s) are represented by a color scale.}
\label{fig:FS}
\end{figure*}

As we explained in the Introduction, due to the transformation of the HT crystal structure to the simple orthorhombic {\it Pbmm} space group, Brillouin zones in all (LT, MT and HT) phases in our calculations have the same shape and high symmetry points (BZ is shown schematically in Fig.~\ref{fig:bands1}). The reciprocal space $x,y,z$ axes are parallel to the $a,b,c$ real space directions. Let us start with some general comments of the band structure, before going into the detailed analysis of the effective masses and the transport properties.
In the LT phase, the highest valence band (VB) appears in the $\Gamma-Z$ direction, parallel to the $z$ axis. It is worth noting, that in the LT phase, this highest VB has a 'pudding mold'--like shape, reported to be beneficial for thermoelectric performance in other systems~\cite{puddingmold}. For the higher temperature structures (MT and HT), valence bands are much different, which points out that electronic band structure is sensitive to temperature, as simulated by the unit cell changes. The 'pudding mold'--like VB changes the shape to become more parabolic, and the second VB at $\Gamma$-Y direction, aligns to band in $\Gamma$-Z direction. The lowest conduction band (CB), seen in the $\Gamma-Y$ direction in all phases, is parabollic-like. 
Above the band gap we see, that the whole band structure in the $\Gamma-Z$ direction considerably moves down, as going from LT to HT, which is correlated with the shortening of the corresponding unit cell $c$ axis. 
In the LT structure, the in-plane unit cell parameters, $b=4.153$~\AA~ and $c=4.445$~\AA~ are significantly different, which in the reciprocal space is reflected in a different alignment of the $\Gamma-Y$ and $\Gamma-Z$ bands. After the phase transition, as well as in the MT case, $b \simeq c$, and the energy locations of CB's minima and VB's maxima become similar between those two directions.
What seems quite surprising, the differences in the band structures between the LT and MT (both before the phase transition) are much larger than between the MT and HT structures (i.e. induced by the phase transition). This shows that the electronic structure evolves continuously with temperature, however the band gap changes abruptly at the phase transition (see, Tab.~\ref{table:tabgap}), which can result in rapid changes in the transport properties.

It is also interesting to analyze the bands in $Z-U$ and $\Gamma-X$ directions, which represent the real-space $a$ direction, i.e. contributing the charge transport between the SnSe layers. In all the cases, for the VB part, the resulting bands are very flat, which is quite intuitive, since we expect hampered charge propagation in this direction. On the other hand, conduction band in $\Gamma-X$ is very steep and almost linear. All these band structure features are reflected in the transport properties of SnSe, as discussed below.

\subsubsection{n-type doping}

\begin{figure}[t]
\centering
\includegraphics[width=0.219\textwidth]{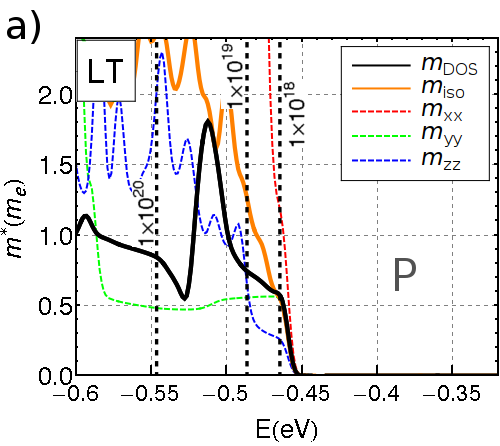}
\includegraphics[width=0.222\textwidth]{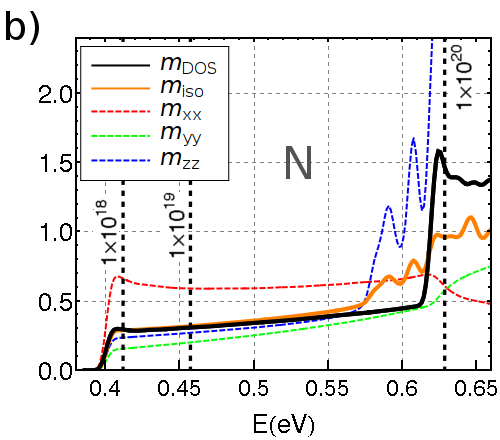}
\includegraphics[width=0.219\textwidth]{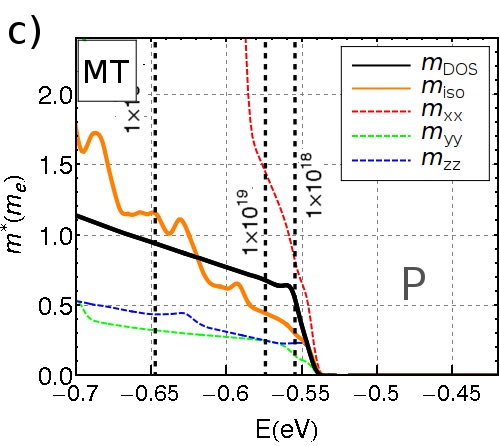}
\includegraphics[width=0.222\textwidth]{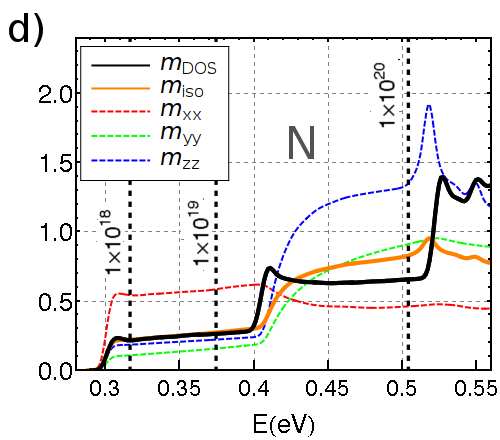}
\includegraphics[width=0.219\textwidth]{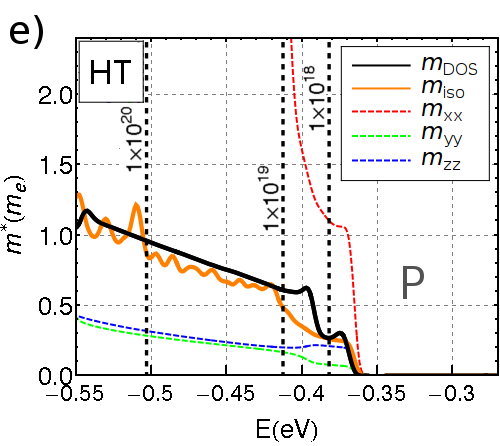}
\includegraphics[width=0.222\textwidth]{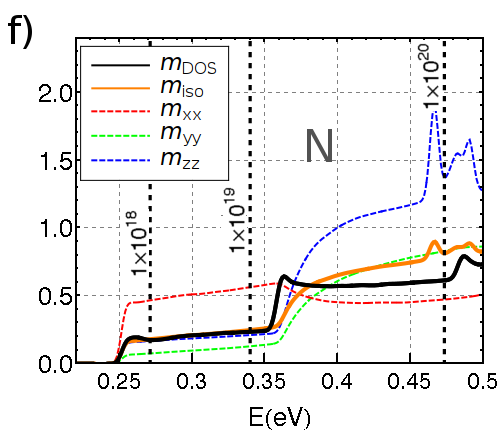}
\caption{(Color online) Effective mass of the LT (a,b), MT (c,d) and HT (e,f) phases of SnSe, calculated using Eq.~\ref{eq:mass} ($m_{\rm DOS}$), Eq.~\ref{eq:mass2} ($m_{xx}$, $m_{yy}$ and $m_{zz}$ ) and Eq.~\ref{eq:massall} ($m_{iso}$). In each case, zero energy is fixed to the middle of the energy gap. Black vertical lines show the Fermi level for various carrier concentrations (in cm$^{-3}$) $1\times10^{18}$, $1\times10^{19}$ and $1\times10^{20}$. Left and right columns present results for the $p$-type and $n$-type doping, respectively.}
\label{fig:m1}
\end{figure}

\begin{figure}[t]
\centering
\includegraphics[width=0.219\textwidth]{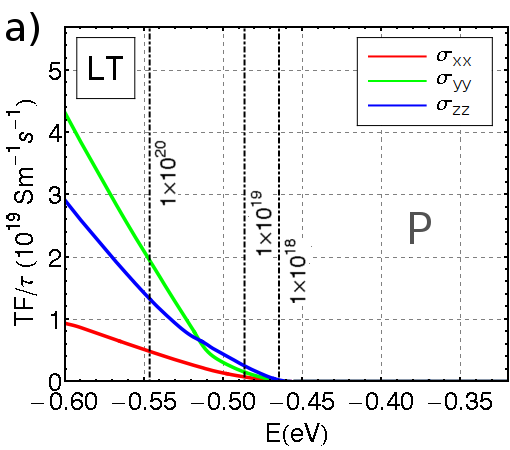}
\includegraphics[width=0.219\textwidth]{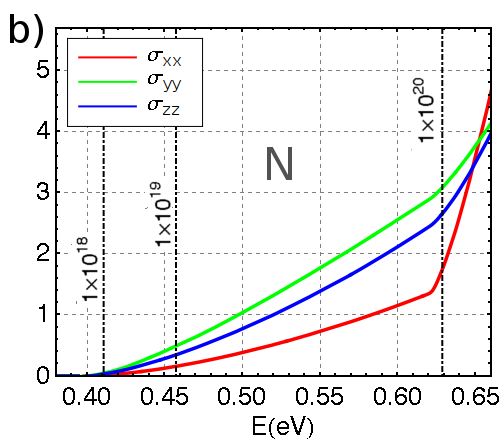}
\includegraphics[width=0.219\textwidth]{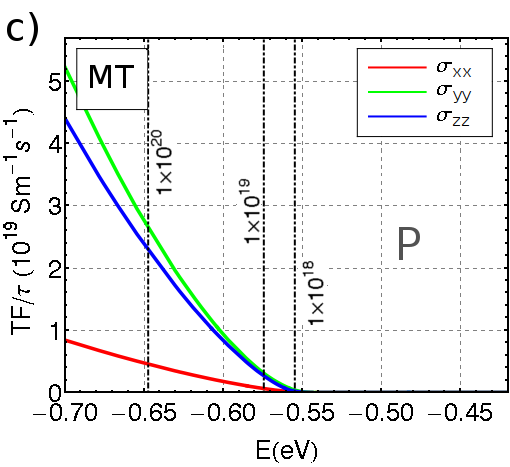}
\includegraphics[width=0.219\textwidth]{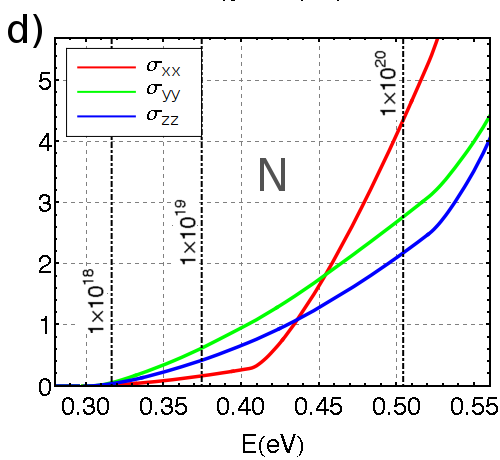}
\includegraphics[width=0.219\textwidth]{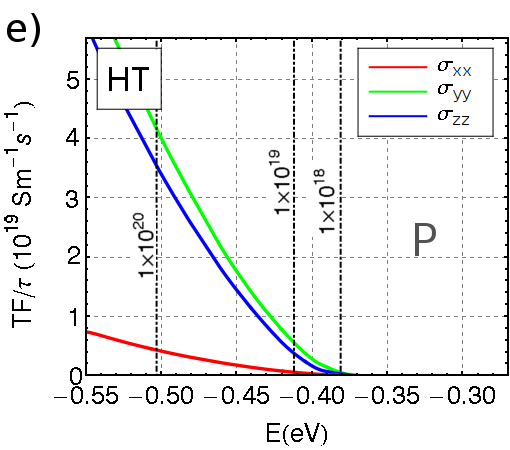}
\includegraphics[width=0.225\textwidth]{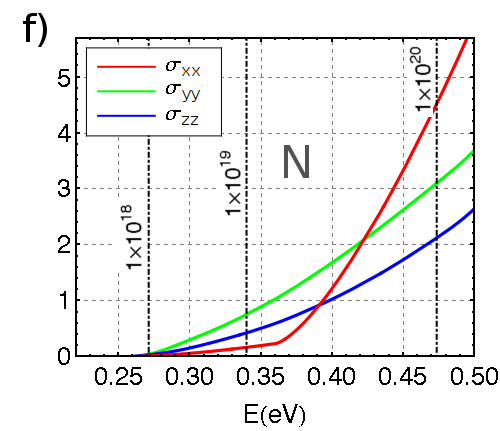}
\caption{(Color online) Transport Function (TF) of valence and conduction bands in LT, MT and HT phases of SnSe. Zero energy is fixed in the middle of energy gap, as in Fig.~\ref{fig:bands1}. Black vertical lines show Fermi level for various carrier concentration (in cm$^{-3}$): $1\times10^{18}$, $1\times10^{19}$ and $1\times10^{20}$. Left and right columns present results for $p$-type and $n$-type doping, respectively.}
\label{fig:TF}
\end{figure}

For the $n$-type doping (whatever the LT, MT and HT structures) at electron concentrations $<10^{19}$~cm$^{-3}$, Fermi level reaches one ellipsoidal electron pocket (with two-fold degeneracy) in $\Gamma$-Y direction (see Fig.~\ref{fig:bands1} and Fig.~\ref{fig:FS} a,b,c). The band, forming this pocket, has a regular, parabolic-like shape, and isotropic effective mass equals $m^*_{iso} \simeq 0.3-0.4$ in LT, and $m^*_{iso} \simeq 0.2-0.3$ in MT and HT case, respectively. The highest values of $m^*$ are found along the $x$-direction (see Fig.~\ref{fig:m1}b,d,f), where they quickly exceed $m^*_{xx} \simeq 0.5$. Therefore, velocity integrated over the Fermi surface (i.e. transport function), is 3 times lower in the $x$-direction than in the $y$- and $z$-directions (see. Fig.~\ref{fig:TF}b,d,f), and low $x$-direction electrical conductivity is expected (since $\sigma_e\sim\sigma(\mathscr{E}_F)$). This behavior corresponds to our expectations of the lower electrical conductivity perpendicular to SnSe layers ($x$-
direction is along $a$-axis).
Electron transport properties change dramatically when the Fermi level reaches further electron pockets; one in HT and MT at the $\Gamma$ point, at $\sim 10^{19}$~cm$^{-3}$, and five in the LT structure (one centered at the $\Gamma$ point and four in the $\Gamma$-T direction) at $\sim 10^{20}$~cm$^{-3}$. At these concentrations, the aforementioned steep linear band in $\Gamma-X$ is activated. In contrast to the previous case, these new pockets have the smallest $m_{xx}$ (see Fig.~\ref{fig:m1}), while the largest mass tensor component is now $m_{zz}$. In consequence, the average effective mass, represented by $m^*_{iso}$ or $m^*_{\rm DOS}$, raises rapidly to $1.0 - 1.5$ (for $n \sim 10^{20}$~cm$^{-3}$ in LT ), and $0.6-0.9$ (for $n \sim 10^{19}$~cm$^{-3}$ in HT and MT). 
Thus, the transport function tensor component $\sigma_{xx}$ (Fig.~\ref{fig:TF}b,d) changes from the smallest, in energy range corresponding to the electron concentration below $10^{19}$~cm$^{-3}$, to the largest, at energies corresponding to $10^{20}$~cm$^{-3}$. The rapid increase of the $\sigma_{xx}$ (connected to the alignment of the five pockets at the same energy) should favor the high thermopower $S_{xx}$, since $S\sim d(\ln\sigma)/d\mathscr{E}$. 
In case of the $n$-type doping, in the LT, MT and HT phases, after reaching certain concentration between $10^{19}$ cm$^{-3}$ (HT) - $10^{20}$ cm$^{-3}$ (LT, see, Fig.~\ref{fig:TF}) two types of electrons are involved in the electrical transport along the $x$-direction: first, those having low mass and high velocity, providing high conductivity, and second, heavy electrons that are needed to achieve high thermopower. In such a case large power factor can be expected. 

DOS effective mass, plotted as a function of the carrier concentration and temperature (temperature effects according to Eq.~\ref{eq:mass_n_T}), is shown in Fig.~\ref{fig:m2}. For a sake of completeness, we present results for all the phases (LT, MT and HT) in the wide $15 - 900$~K temperature range.
Appearance of the additional electron pockets, that is manifested in the rapid rise of the effective mass, have sharply determined concentrations only at very low temperatures. For $T=300$~K, due to temperature blurring of the Fermi-Dirac statistics, $m^*$ changes gradually with a pronounced bump above $10^{20}$~cm$^{-3}$ in the LT structure. At temperature c.a. $T=800$~K or higher, electrons, which actively conduct, presumably come from the deeper lying bands, that have strongly non-parabolic dispersion relations. Thus, effective mass is not well defined, and the results must be treated with caution. Generally, in all the phases, DOS effective masses rise with the concentration, from approximately $0.3$~$m_e$ at $n=10^{18}$~cm$^{-3}$ to $2-3$~$m_e$ at $n=10^{21}$~cm$^{-3}$.   

\subsubsection{p-type doping}

\begin{figure}[b!]
\centering
\includegraphics[width=0.49\textwidth]{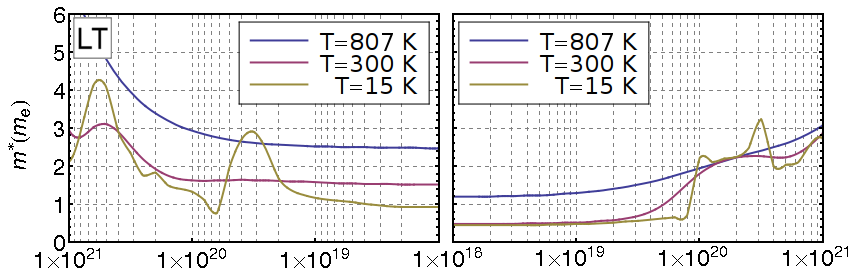}
\includegraphics[width=0.49\textwidth]{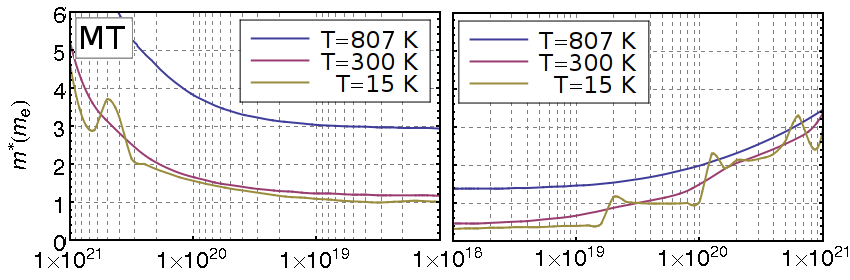}
\includegraphics[width=0.49\textwidth]{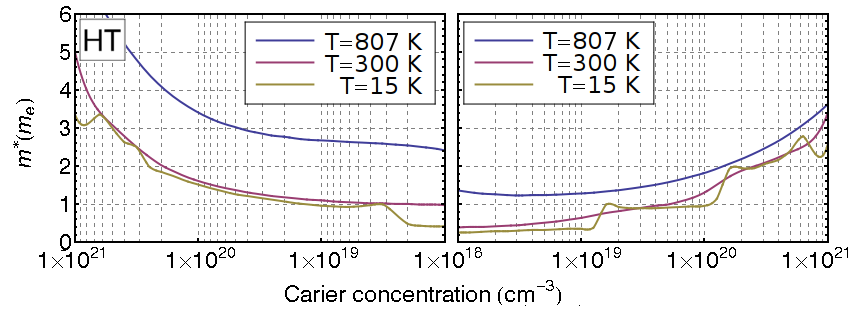}
\caption{(Color online) DOS effective mass in function of the carrier concentration and temperature for the LT, MT and HT phases of SnSe. Left and right columns correspond to the $p$-type and $n$-type doping, respectively.}
\label{fig:m2}
\end{figure} 
In the $p$-type LT-structure, up to 10$^{20}$~cm$^{-3}$, the Fermi level penetrates hole pockets located in $\Gamma$-Z direction (Fig.~\ref{fig:FS}d). These pockets have non-ellipsoidal shape, which is caused by the 'pudding-mold'-like band, having two maxima (see Fig.~\ref{fig:bands1}a). In this case, effective mass cannot be well defined (see Fig.~\ref{fig:m2}a), since such a band exhibits both, electron-like, and hole-like properties, due to the convex and concave curvatures. However, the Fermi surface shape indicates, that the $x$-direction effective mass possesses the highest value (elongated shape in $x$-direction), which, in real space, corresponds to the $a$-axis direction (perpendicular to the SnSe layers).
This behavior obviously affects the carrier velocity, integrated over the Fermi surface (see, Fig.~\ref{fig:TF}a). The component $\sigma_{xx}$, similar to $n$-type case, is again the lowest and the highest values of TF are detected in $z$- (close to VB edge) and $y$-directions (well below VB edge), both being parallel to the SnSe atomic layers. Interestingly, at  higher temperatures (MT and HT phases) the bands in $\Gamma$-Z direction do not have inflection, and the effective mass is therefore well defined.
At higher concentrations (10$^{20}-10^{21}$~cm$^{-3}$) in LT , the Fermi level also reaches holes from the $\Gamma$-Y direction (see Fig.~\ref{fig:bands1}a and Fig.~\ref{fig:FS}a). 
In HT (and also MT), the axial anisotropy of the transport properties of SnSe ($m^*$ and TF) is nicely visible: very large effective mass and the smallest transport function is seen in $x$-direction, compared to much more 'conducting' $y$- and $z$-directions (in-plane transport). 
The electrical conductivity $\sigma_{e,xx}$ is expected to be about three times lower (see Fig.~\ref{fig:TF}c) than in the other directions. 

Figs.~\ref{fig:m2} (left column) show $p$-type DOS effective mass in function of the carrier concentration and temperature. The LT structure, already at $n=10^{18}$~cm$^{-3}$, has considerably large value of the 'bare' mass (i.e. not affected by temperature blurring, $T=15$~K curve), $m^* \simeq 1.0$, which is raising significantly above $n=2\times10^{19}$~cm$^{-3}$.  
At $T=300$~K, similarly to the $n$-type doping, critical concentration, where additional bands start to have influence on $m^*$, is blurred, and $m^*$ is almost constant ($m^* \simeq 1.5$) up to $10^{20}$~cm$^{-3}$, then rising to $3$~$m_e$ at $n=10^{21}$~cm$^{-3}$. 
At higher temperatures, the deep and heavy valence bands contribute to the effective mass even at the lowest concentrations, however, we have to keep in mind, that integral in Eq.\ref{eq:mass_n_T} covers carriers being far from the gap, where bands are generally not parabolic, thus the characterization of bands in terms of the effective mass may become inaccurate. 

Closing the effective mass discussion, for all the LT,MT and HT cases, $n$-type effective masses are smaller than $p$-type.

\subsection{Transport properties}
\label{sec:thermo}

\subsubsection{Thermopower}

\begin{figure}[b]
\centering
\includegraphics[width=0.499\textwidth]{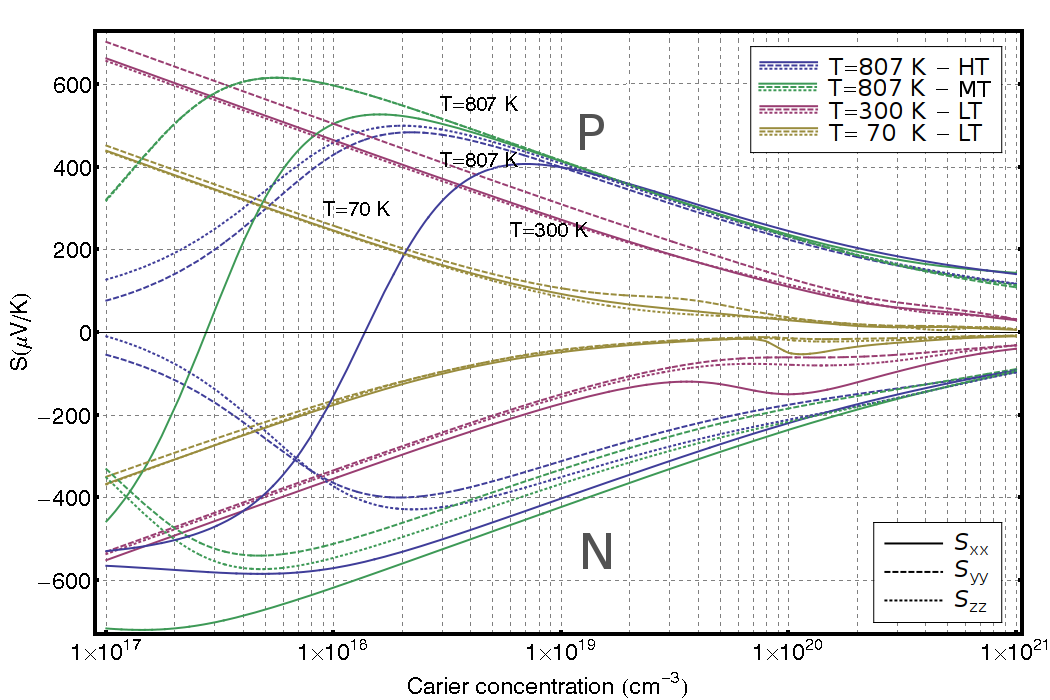}
\caption{(Color online) Thermopower, in function of the carrier concentration, for the $p$- (top panel) and $n$-type (bottom panel) SnSe, for $T = 70$~K, 300~K and 807~K. The electronic band structures of different phases were used to calculate thermopower: LT at 70 K and 300 K,  MT and HT at 807 K.}
\label{fig:S_n}
\end{figure}
Thermopower as a function of electrons/holes concentration for the three different temperatures are presented in Fig.~\ref{fig:S_n} where data for the different temperatures were obtained for the corresponding crystal structures. 
The complete results for all the phases at all the temperatures are shown in the Appendix~\ref{sec:append2}. 
The most interesting features are seen for the $n$-type LT case, that exhibits strong anisotropy of the Seebeck coefficient, visible at $T=70$ and $300$~K around $n \sim 10^{20}$~cm$^{-3}$. The $n$-type $S_{xx}$ tensor element has the largest absolute value; at low temperatures $S_{xx}$ starts to dominate at $1\times10^{20}$~cm$^{-3}$, where the Fermi energy reaches 5 electron pockets with high $v_x$ (see, red line in Fig.\ref{fig:TF}b). 
At higher temperatures (i.e. for MT and HT), around $n \sim 2\times10^{19}$~cm$^{-3}$, $E_F$ reaches only one additional electron pocket (see, Fig.~\ref{fig:TF}d,f, red line), and the advantage of reaching this pocket spreads over wider concentrations range (see, $n$-type $S_{xx}$ at 807~K in Fig.~\ref{fig:S_n}). This is reflected as $\sim20\%$ rise of $S_{xx}$ thermopower in the whole concentrations range.

The anisotropy of $p$-type thermopower is less significant.
Similar enhancement of the thermopower, as observed for the $n$-type, caused by the appearance of additional bands near the Fermi level, is seen in $p$-type $S_{yy}$ component at $p\approx3\times10^{19}$~cm$^{-3}$. Two, non-parabolic pockets, have a high velocity of electrons in $y$-direction, increasing rapidly at $p\approx3\times10^{19}$~cm$^{-3}$, which provides the increase of $\sigma_{yy}$ derivative (see, green line in Fig.~\ref{fig:TF}a). 
Significance of this effect is smaller, than in the $n$-type doping, and seen only in the LT phase, but it is sufficient to provide an increase of $S_{yy}$ over $S_{xx}$ and $S_{zz}$ (best seen at 300 K).

As far as the bipolar effects are concerned, we can observe, that $p$-type $S_{xx}$ element shows the strongest reduction at the high temperature, which is caused by the large $n$-type $S_{xx}$ and $\sigma_{xx}$. 
The major detrimental effect at high temperatures is related to the reduction of the band gap during the phase transition (LT and MT vs HT, see Table~\ref{table:tabgap}). To visualize this influence of the gap reduction on the bipolar effect, we have plotted $S$ at the same temperature of 807 K (phase transition temperature) computed for the MT and HT cases. Much smaller band gap in the HT phase leads to the much stronger bipolar effect, and the thermopower starts to decrease already for $p < 10^{19}$ cm$^{-3}$ and $n < 2\times10^{18}$ cm$^{-3}$. 
Note also, that the bipolar effect critically depends on the value of the band gap, which here for the HT phase was adjusted in an approximate way, since we are not aware of the experimental value. 

\begin{figure}[!t]
\centering
\includegraphics[width=0.249\textwidth]{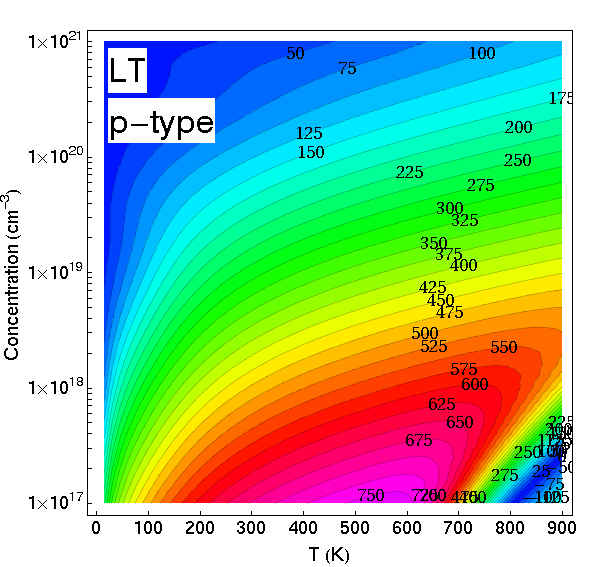}
\includegraphics[width=0.228\textwidth]{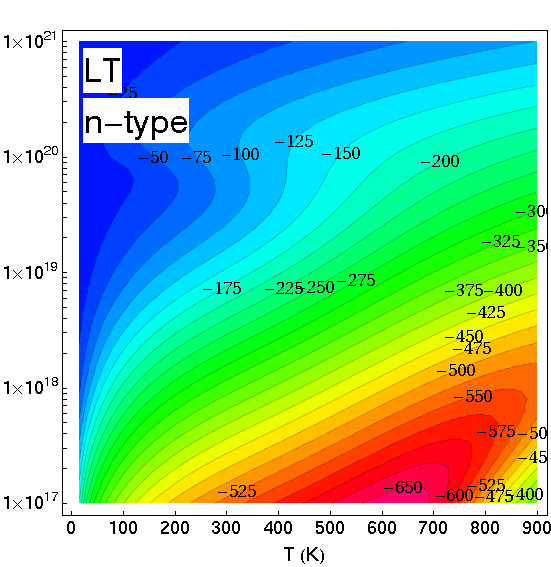}
\includegraphics[width=0.249\textwidth]{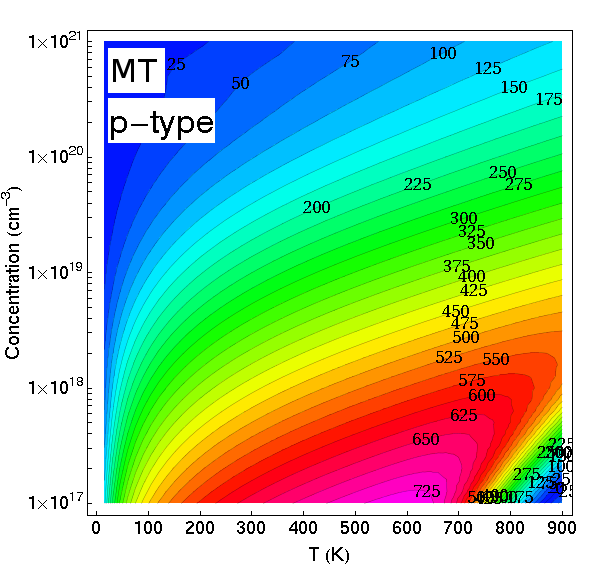}
\includegraphics[width=0.228\textwidth]{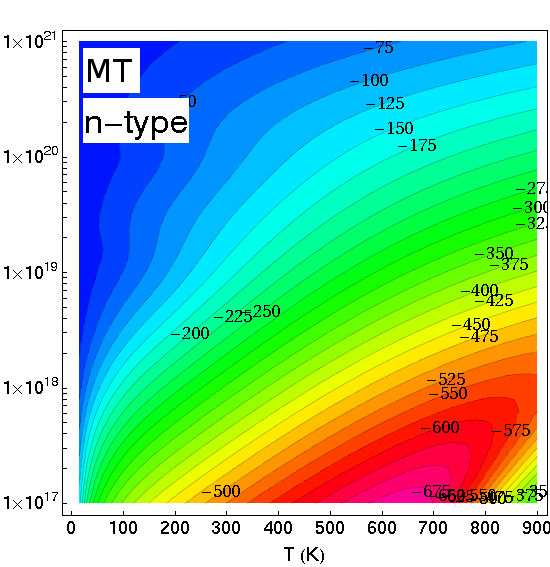}
\includegraphics[width=0.249\textwidth]{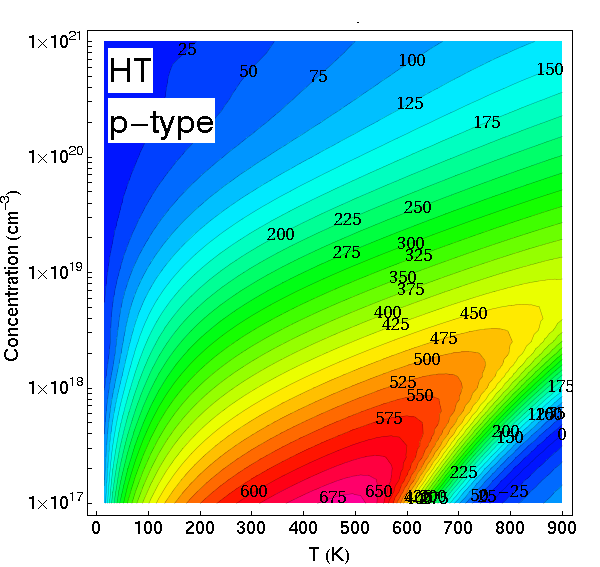}
\includegraphics[width=0.228\textwidth]{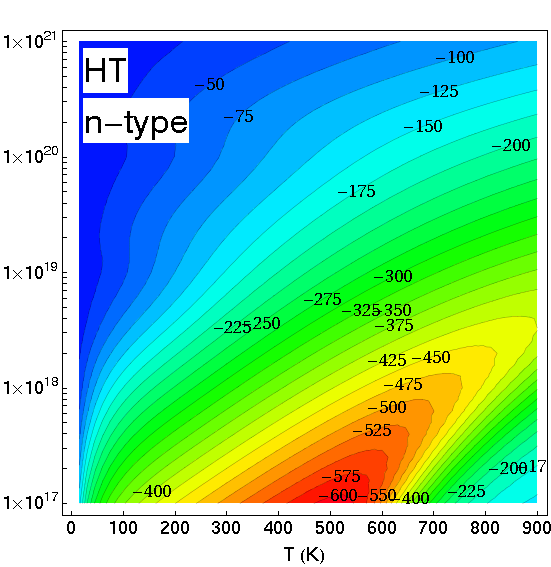}
\caption{(Color online) Isotropic thermopower for LT, MT, and HT phases of SnSe (see Tab.~\ref{table:structure}) for $n$- and $p$-type doping.}
\label{fig:Siso}
\end{figure}

Isotropic thermopower, calculated using Eq.~\ref{eq:Save}, is shown as a color map in Fig.~\ref{fig:Siso}, and $S(T)$ curves for selected carrier concentrations are collected in Appendix~\ref{sec:append2}, Fig.~\ref{fig:ST}. 
To allow for a comparison between the different phases, all three cases for complete temperature ranges are displayed.
After averaging over three directions, in all the three phases, $p$-type thermopower is generally larger than $n$-type one. 
For $n$-type LT map, around $n \sim 10^{20}$~cm${-3}$, and below 400~K, we observe the abnormal bending of $S$, which increases with the carrier concentration. This is due to the rise in $S_{xx}$ element, discussed before. For the MT and HT phases, such effect is not observed, and the variation of $S$ with the carrier concentration is monotonic ($S$ decreases with $n$ or $p$).
At the highest temperatures and the lowest carrier concentrations, we again observe a drop of $S$ due to the bipolar effects, the strongest for the $p$-type HT phase.

\subsubsection{Power factor}

The power factor (PF$=S^2/\sigma$), similarly to the electrical conductivity, cannot be directly calculated if the electronic relaxation time $\tau$ is not known, however the discussion of PF$/\tau$ can be still very useful in optimizing the carrier concentration in the SnSe system.
For completeness of the analysis, we show results for all the  phases (LT, MT and HT) at $T = 450$ and $807$~K, although one has to bear in mind that these phases describe SnSe in limited temperature ranges. 

\begin{figure}[htb]
\centering
\includegraphics[width=0.24\textwidth]{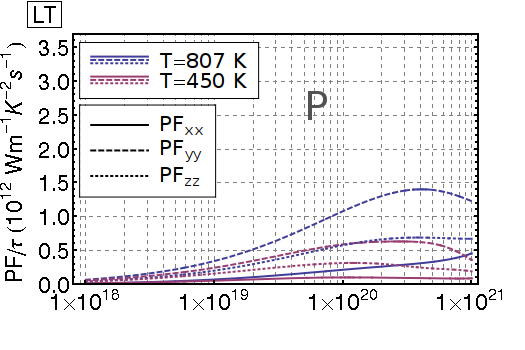}\includegraphics[width=0.24\textwidth]{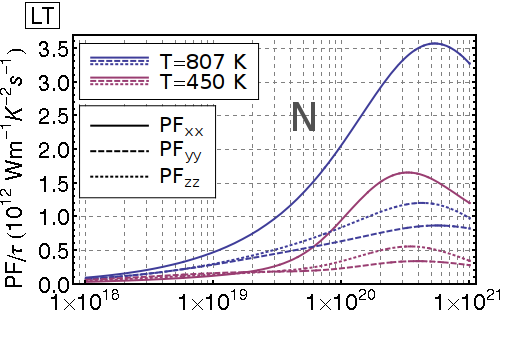}
\includegraphics[width=0.24\textwidth]{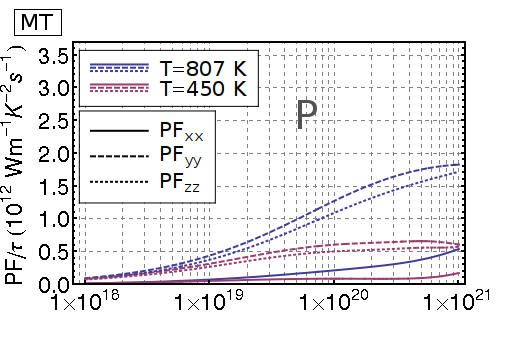}\includegraphics[width=0.24\textwidth]{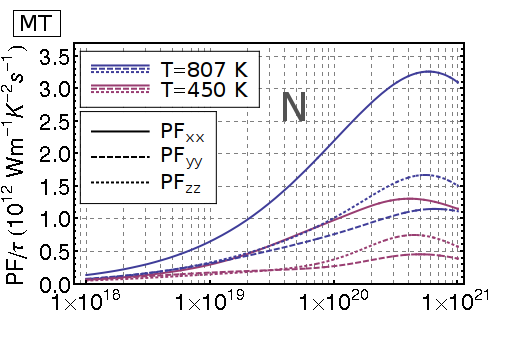}
\includegraphics[width=0.24\textwidth]{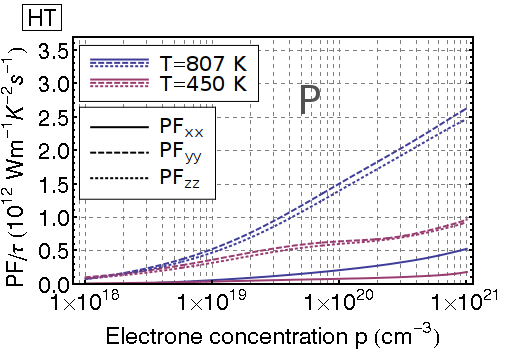}\includegraphics[width=0.24\textwidth]{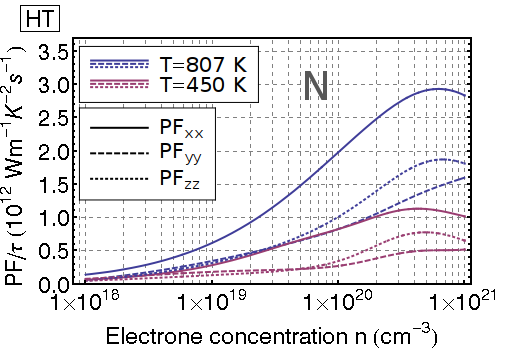}
\caption{(Color online) Power factor of the LT, MT and HT phases for $p$- (left column) and $n$-type (right column) SnSe in function of concentration for two temperatures (450 K and 807 K), calculated along three directions.  }
\label{fig:PF1}
\end{figure}

\begin{figure}[hbt]
\centering
\includegraphics[width=0.249\textwidth]{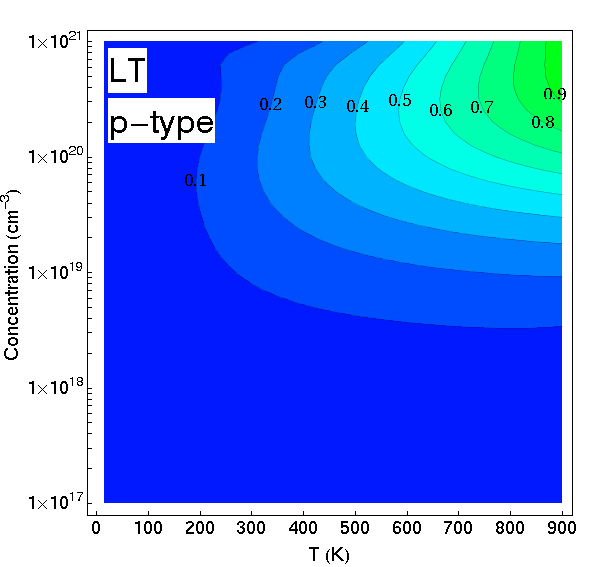}
\includegraphics[width=0.228\textwidth]{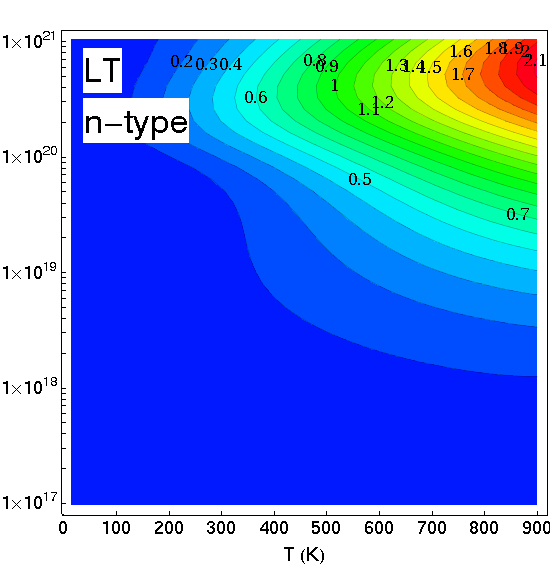}
\includegraphics[width=0.249\textwidth]{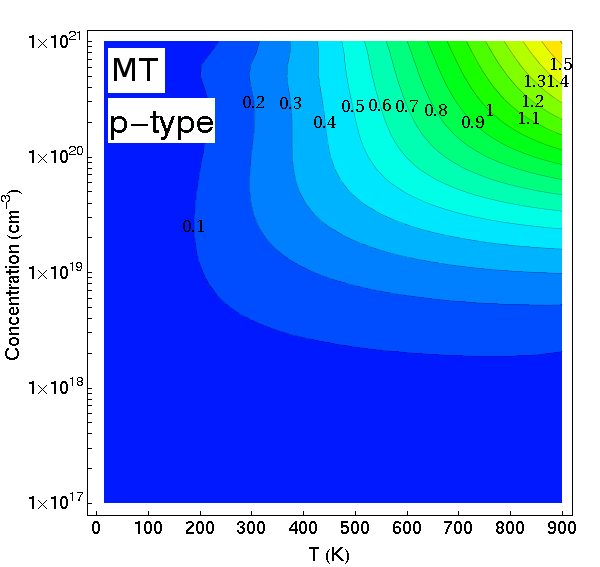}
\includegraphics[width=0.228\textwidth]{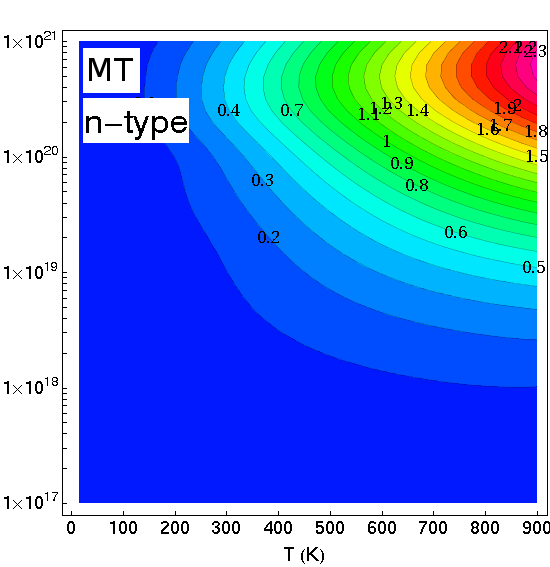}
\includegraphics[width=0.249\textwidth]{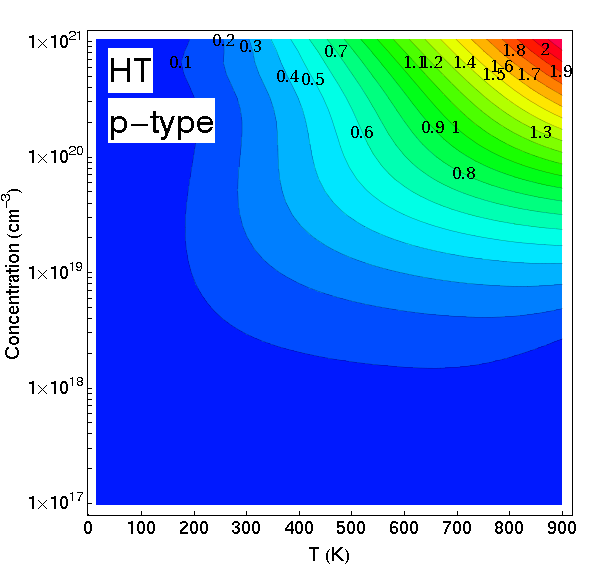}
\includegraphics[width=0.228\textwidth]{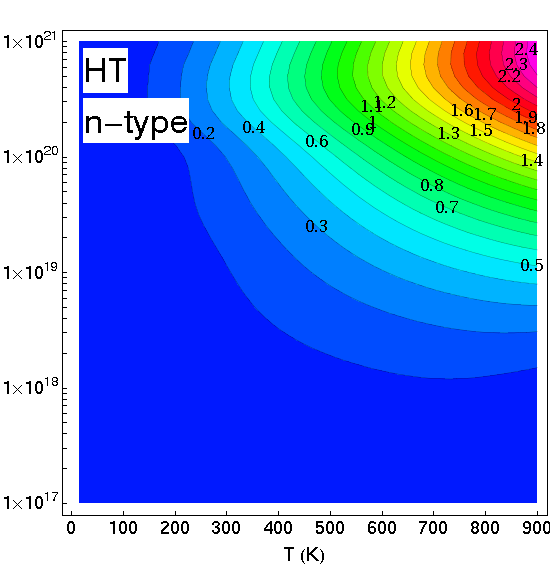}
\caption{(Color online) Isotropic power factor for $n$- and $p$-type SnSe (LT, MT and HT phases) in function of the carrier concentration and temperature. All values in $10^{12} WK^{-2}m^{-1}s^{-1}$. }
\label{fig:PF2}
\end{figure}

Results of the calculations are shown in Fig.~\ref{fig:PF1}. The most striking observation is, that for  the  $p$-type SnSe, $xx$ tensor element of the power factor (PF$_{xx}$) is the smallest one, in contrast to the $n$-type, where PF$_{xx}$ is the largest. The reason is that in the  $p$-type, $x$-direction inter-layer conducting channel is blocked due to the smallest TF (see, Fig.~\ref{fig:TF}, red line), whereas in the  $n$-type, it is activated by the electron pockets with high $v_x$.
The smallest $p$-type PF$_{xx}$, comparing to $y$- and $z$-directions, remains in agreement with the experimental findings~\cite{SnSeNature}.

In the  $p$-type LT phase, value of  the thermopower is approximately isotropic (except for $S_{xx}$ at low concentrations), thus overall anisotropy in the  PF is due to the  electrical conductivity: transport function element $\sigma_{yy}/\tau$ has the biggest value (in $p>5\times 10^{19}$~cm$^{-3}$) which makes PF the largest in the  $y$-direction.  
For the  high temperature structures, namely MT and HT, the power factor for $y$- and $z$-directions becomes similar, and is the largest for the  HT phase. 
In  the $n$-type SnSe, PF$/\tau$ curves look quite similar, while comparing the  LT, MT and HT cases, with decreasing anisotropy (measured as PF$_{xx}$/PF$_{yy}$ and PF$_{xx}$/PF$_{zz}$) upon going from LT to HT.

For isotropic (polycrystalline) material,  the power factor, calculated using Eq.~\ref{eq:PFave}, is mapped in  Fig.~\ref{fig:PF2}.
The first observation is, that generally  the $n$-type SnSe has larger values of PFs, than the  $p$-type one, thus the heavily doped $n$-type SnSe can exhibit even better thermoelectric performance, and better polycrystalline $zT$ over the broad temperature range.
It seems that the existence of the structural phase transition is beneficial for the power factor value only for $p$-type SnSe, where HT phase exhibits the largest values, but only for large carrier concentrations, where the bipolar effects are not so important. However, at higher concentrations, the $zT$ value may not benefit from the phase transition, because of the increasing contribution of the electronic part of the thermal conductivity.

The maps of PF also reveal that SnSe is best suited for TE application at temperatures $T > 600$~K and should be rather heavily doped, for both $n$- and $p$-type above $10^{20}$~cm$^{-3}$.

Note, that all the presented transport calculations results and predictions are based on the constant relaxation time approximation, and, in principle, electron-phonon or electron-impurity scattering may markedly affect them. Especially for the case, where both heavy, and light electrons are involved in electrical transport, scattering effects may be of increased importance, due to the different scattering rates for these two electronic bands.

\subsection{Comparison with experiment}

\begin{figure}[ht]
\centering
\includegraphics[width=0.48\textwidth]{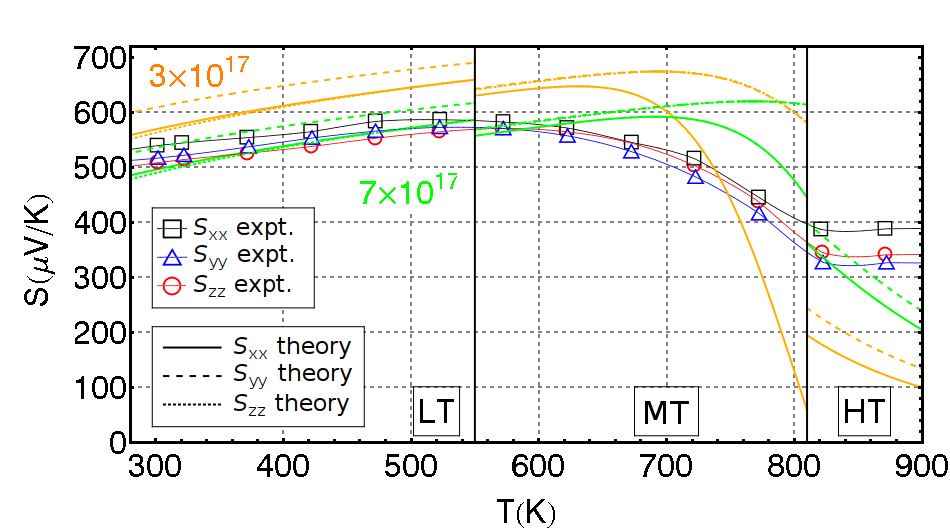}
\caption{(Color online) Comparison of  the  calculated and experimental values~\cite{SnSeNature} of  the thermopower vs. temperature. Nominal hole concentrations for  the theoretical curves are $3\times10^{17}$ (orange) and $7\times10^{17}$~cm$^{-3}$ (green).}
\label{fig:expcalcS}
\end{figure}

\begin{figure}[hbt]
\centering
\includegraphics[width=0.23\textwidth]{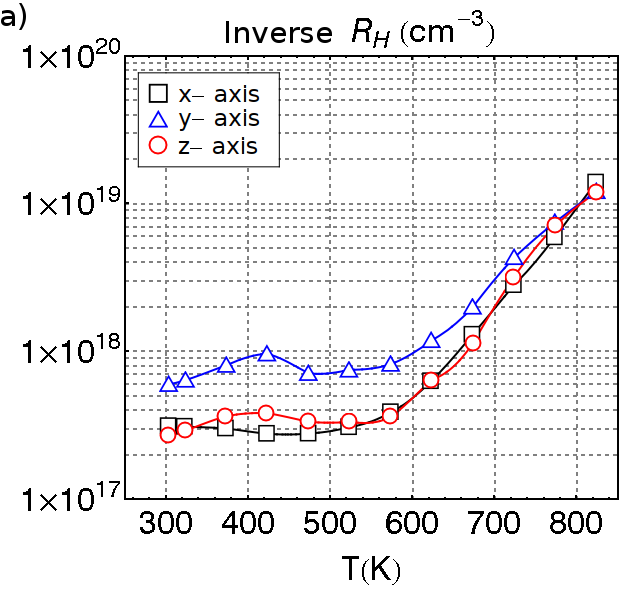}
\includegraphics[width=0.23\textwidth]{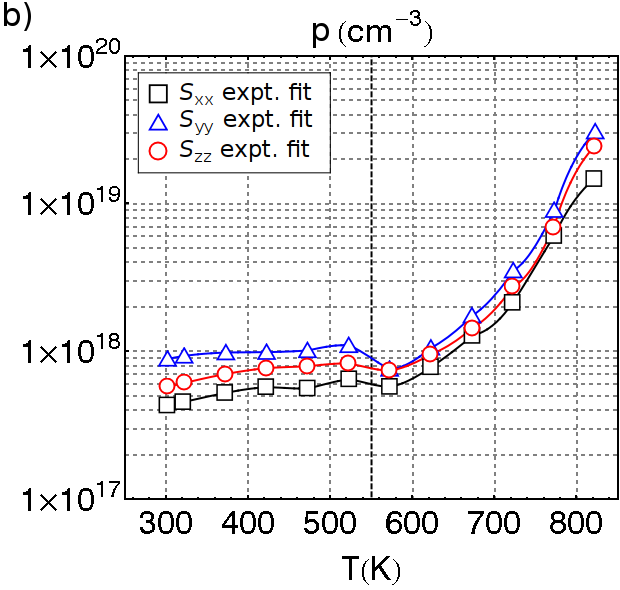}
\caption{(Color online) (a) Concentration derived from  the measured Hall coefficient, according to $n_H=1/eR_H$); (b) nominal carrier concentration at which  the calculated thermopower equals to the experimental value. All concentrations are given in cm$^{-3}$, experimental data are taken from Ref.~\onlinecite{SnSeNature}. Vertical line separates the temperature ranges, where theoretical thermopower of LT (below $550$~K) and MT (above $550$~K) phases was used. }
\label{fig:ConFit}
\end{figure}

Experimental data were taken from Ref.~\onlinecite{SnSeNature}, where the transport properties of the  single crystal $p$-type SnSe were reported. Since our calculations were performed within the constant relaxation time approximation, only thermopower can be directly compared. Fig.~\ref{fig:expcalcS} shows the measured Seebeck coefficient along $a$, $b$, and $c$ crystallographic directions\footnote{The original data from Ref.~\onlinecite{SnSeNature} were transformed to our convention of the unit cell axes. In Ref.\onlinecite{SnSeNature} axes $c$ is the shortest one, thus $b$ and $c$ are exchanged, comparing to our convention of Pnma.} 
compared with two sets of corresponding theoretical values, for hole concentrations $p=3\times10^{17}$ and $7\times10^{17}$~cm$^{-3}$. 
These carrier concentrations were selected for comparison basing on the measured Hall coefficient of SnSe at $T=300$~K, depicted also in Fig.~\ref{fig:ConFit}a. Nevertheless, one has to bear in mind that due to the anisotropic crystal structure and Fermi surface, as well as the strongly non-parabolic band structure, Hall coefficient data do not reflect directly the carrier concentration.
In Fig.~\ref{fig:expcalcS} we see, that the experimental Seebeck coefficient well matches the theoretical values below 600~K, for $p=7\times10^{17}$ ~cm$^{-3}$, and then is underestimated in the mid-temperature range ($\sim 700$~K). The most intriguing experimental behavior is noticed for the thermopower, which remains constant with temperature above $T_c$. For such high temperatures, and carrier concentration around $p\sim 10^{17}$~cm$^{-3}$, bipolar effect should strongly decrease $S$, even making $S$ negative, as seen from our computations, shown in Fig.~\ref{fig:S_n} and Fig.~\ref{fig:ST}. Similar observations, with even faster decreasing $S$ due to the smaller band gap, were reported from calculations in Ref.~\onlinecite{snse-arxiv}.

One can attempt to explain such an uncommon situation assuming the gradual increase of  the carrier concentration with temperature, since it could explain both, the decrease in $S$ above 600~K, and saturation of $S$ above 800~K, with almost invisible bipolar effect.
Experimental Hall measurements also show, that $1/R_H$ rises almost two order of magnitudes (see, Fig.~\ref{fig:ConFit}a), between 600 and 800~K, which supports the above-mentioned hypothesis. 
Moreover, for  the high carrier concentration, the thermopower changes smoothly while crossing the phase transition temperature. Fig.~\ref{fig:S_n} shows that only concentrations larger than $1\times10^{19}$~cm$^{-3}$ provide similar Seebeck coefficient at $T=807$~K for MT and HT phases. 
Finally, measured electrical conductivity (see, Ref.~\onlinecite{SnSeNature}) shows non-monotonic behavior with increasing temperature, i.e. it decreases from 300 K to 550 K, increases above 550 K and becomes constant above $T_c$. Such a trend remains in line with the expectation of increasing carrier concentration above 550 K, where the experimental and theoretical thermopower starts to deviate. 

To partly verify this hypothesis, we have plotted how carrier concentration should change with temperature, to reach the agreement between the experimental and calculated thermopowers.\footnote{Carrier concentration $p$ is extracted, applying the condition, that $S_{\rm expt.}(T) = S_{\rm theor.}(p,T)$ at every $T$.} 
As we can see in Fig.~\ref{fig:ConFit}b, the unusual, experimental behavior of thermopower, directly results in the rising carrier concentration in SnSe, to more than $1\times10^{19}$~cm$^{-3}$ around 800~K, and the measured inverse Hall resistivity (Fig.~\ref{fig:ConFit}a), exhibits similar temperature dependence to our extracted $p(T)$ function.

\section{Summary}
\label{sec:Conclusions}

Electronic structure and thermoelectric properties of $n$ and $p$-type SnSe were studied, using the KKR method and the  Boltzmann transport approach, within the constant relaxation time approximation.  
We have shown, that the temperature evolution of the crystal structure within the $Pnma$ phase (i.e. before reaching the structural phase transition at 807 K), leads to  the  significant changes in the electronic band structure.
On the other hand, the phase transition, occurring at $807$~K, leads mainly to the abrupt change of the energy band gap value, whereas modifications of $\mathscr{E}({\bf k})$ curves are minor. 
The effective masses, analyzed in function of the carrier concentration and temperature, indicate, that overall $p$-type masses are larger than $n$-type ones, and SnSe exhibits strong anisotropy of the electron transport properties for both types of charge conductivity. 

$P$-type SnSe, computed for the room temperature crystal structure parameters, has strongly non-parabolic dispersion relations, with the 'pudding-mold'-like shape of the highest valence band. Due to the flat bands, yielding large effective masses, the inter-layer electron transport seems to be blocked, resulting in small power factor and Seebeck coefficients, when comparing with the corresponding values computed for the other directions (i.e. in-plane electron transport). 

The opposite situation was found in the $n$-type SnSe, where the inter-layer direction exhibits band structure features, highly beneficial for thermoelectric performance. In this case, when critical concentration between $10^{19}$ cm$^{-3}$ (HT) - $10^{20}$ cm$^{-3}$ (LT, see, Fig.~\ref{fig:TF}) is reached, or when high velocity electron band is thermally activated at high temperatures, two types of carriers are present simultaneously in the system, i.e. more localized electrons, having large effective masses, and highly mobile electrons, possessing low effective masses and high velocity. 
This particular combination results in high thermopower and power factor in the inter-layer direction. In view of our results,  the $n$-type SnSe may be even better thermoelectric material, than the $p$-type one. 

Our theoretical study confirmed the strong anisotropy of the electron transport in SnSe. This results in the much worse thermoelectric performance in the polycrystalline material, in agreement with the experimental reports~\cite{snse-Lenoir,snse-Snyder}.
 
On the whole, the computational results, derived from the KKR method and Boltzmann equations, show quite good agreement with  the measured data~\cite{SnSeNature} below $T \sim 600$~K. The experimentally observed changes in $S$, $\sigma$ and $R_H$ for $T > 600$~K suggest, that in the measured SnSe samples, the carrier concentration was gradually increasing, and the discrepancy between the theoretical and measured Seebeck coefficient, appearing at high temperatures ($T> 600$~K) can be corrected when  the actual Hall concentration is taken into account in calculations. 
Nevertheless, the mechanism responsible for such a generation of additional carriers, suggested in our discussion, remains unknown.

\appendix
\section{Polycrystalline random grain model}
\label{sec:append1}
\begin{figure}[hbt]
\centering
\includegraphics[width=0.26\textwidth]{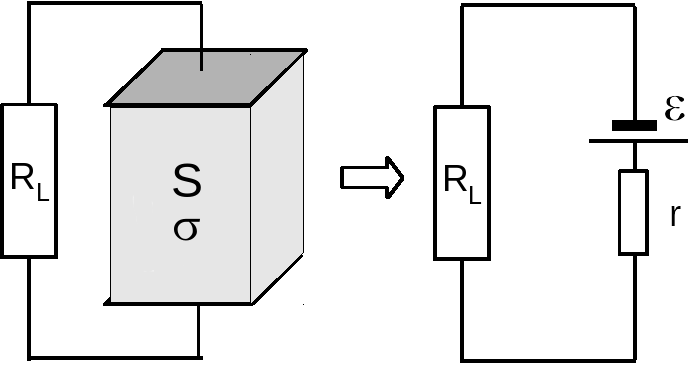}
\caption{Circuit model of lumped (with $R_L$) thermoelectric material.}
\label{fig:app1}
\end{figure}

In a polycrystalline material, where grains have different crystalline orientation and are randomly distributed, the effective Seebeck coefficient, as well as  the power factor, are not simple averages of the single grain properties. To obtain the correct isotropic values, appropriate circuit model for thermoelectric generator is build (see Fig.~\ref{fig:app1}).

\begin{figure}[hbt]
\centering
\includegraphics[width=0.49\textwidth]{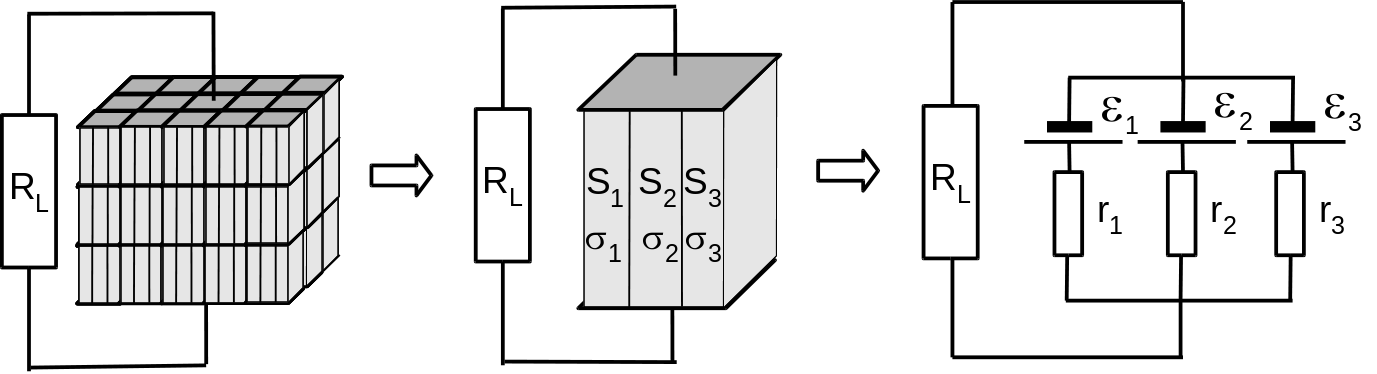}
\caption{Model of  the  thermoelectric material, which is build up from three types of small, randomly distributed grains, with different transport properties.}
\label{fig:app2}
\end{figure}

In  the polycrystalline sample it is assumed, that the material consists of three types of small grains, with thermopower $S_k$ ($k=1,2,3$) each, and electrical conductivities $\sigma_k$ (see Fig.~\ref{fig:app2}), distributed on a regular grid. This model can be simplified to the thermoelectric module made from three solid materials connected in parallel. On this basis, equivalent circuit model was made.

The thermoelectric module with one  the  type of material, having  the  thermopower $S$ and electrical conductivity $\sigma$, can be replaced with  the electromotive force (emf) $\mathcal{E}=S\Delta T$ and $r=\sigma l/A$, where $l$ and $A$ are length, and cross section of the module, respectively (see Fig.~\ref{fig:app1}). Effective thermopower can be defined as voltage on the module, divided by the temperature difference $\Delta T$, when $R_L$ goes to infinity. 

\begin{equation}
S_{eff}=\frac{U}{\Delta T}=\frac{iR_L}{\Delta T}=\frac{\mathcal{E}}{r+R_L}\frac{R_L}{\Delta T}\xrightarrow{R_L\rightarrow \infty}\frac{\mathcal{E}}{\Delta T}=S
\end{equation}
where $i$ is the current in circuit. 

The power factor can be defined as a capability of energy production of material with the cross section $A$, and length $l$, at the given temperature difference $\Delta T$
\begin{equation}
PF_{eff}=\frac{Pl}{A\Delta T^2},
\end{equation}
where $P$ is the power output of the source, when $R_L$ goes to zero (short circuit power).
\begin{equation}
P=\mathcal{E}i=\frac{\mathcal{E}^2}{r+R_L}\xrightarrow{R_L\rightarrow 0}\frac{\mathcal{E}^2}{r}
\end{equation}
and therefore
\begin{equation}
PF_{eff}=\frac{Pl}{A\Delta T^2}=\left(\frac{\mathcal{E}}{\Delta T}\right)^2\frac{l}{A r}=S^2\sigma,
\end{equation} 

In case of  the equivalent polycrystalline circuit model,  the  effective thermopower is:
\begin{equation}
S_{eff}=\frac{U}{\Delta T}=\frac{iR_L}{\Delta T}.
\end{equation}
From circuit theory
\begin{equation}
i_k=\frac{\mathcal{E}_k-iR_L}{r_k}\text{ \ and \ \ } i=\sum_{k}i_k
\end{equation}
where $k$ is $1,2,3$. Now $i$ can be found
\begin{equation}
i=\frac{\sum_k \mathcal{E}_k/r_k}{R_L\sum_k 1/r_k+1}.
\end{equation}
\begin{equation}
\begin{split}
S_{eff}&=\frac{iR_L}{\Delta T}=\frac{\sum_k \mathcal{E}_k/r_k}{R_L\sum_k 1/r_k+1/R_L}\frac{1}{\Delta T}\xrightarrow{R_L\rightarrow \infty}\\=&\frac{\sum_k \mathcal{E}_k/r_k}{\sum_k 1/r_k}\frac{1}{\Delta T}=\frac{\sum_k S_k\sigma_k}{\sum_k \sigma_k}.
\end{split}
\end{equation}
Power of the source is now a sum of powers of all the sources
\begin{equation}
P=\sum_k\mathcal{E}_ki=\frac{\mathcal{E}^2_k-i\mathcal{E}_kR_L}{r_k}=\xrightarrow{R_L\rightarrow 0}=\sum_k\frac{\mathcal{E}^2_k}{r_k}
\end{equation}
and therefore
\begin{equation}
PF_{eff}=\frac{Pl}{A\Delta T^2}=\sum_k\left(\frac{\mathcal{E}_k}{\Delta T}\right)^2\frac{l}{A_k r_k}=\frac{1}{3}\sum_kS_k^2\sigma_k,
\end{equation}
where $A_k=1/3A$ (grains are randomly distributed).

\section{Extended data}
\label{sec:append2}
\begin{figure}[H]
\centering
\includegraphics[width=0.499\textwidth]{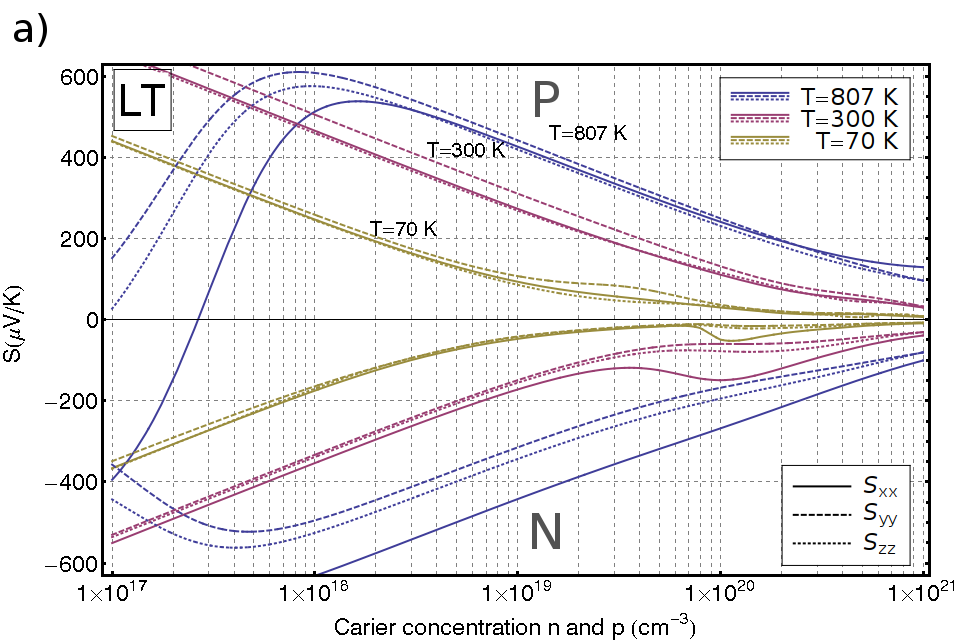}
\includegraphics[width=0.499\textwidth]{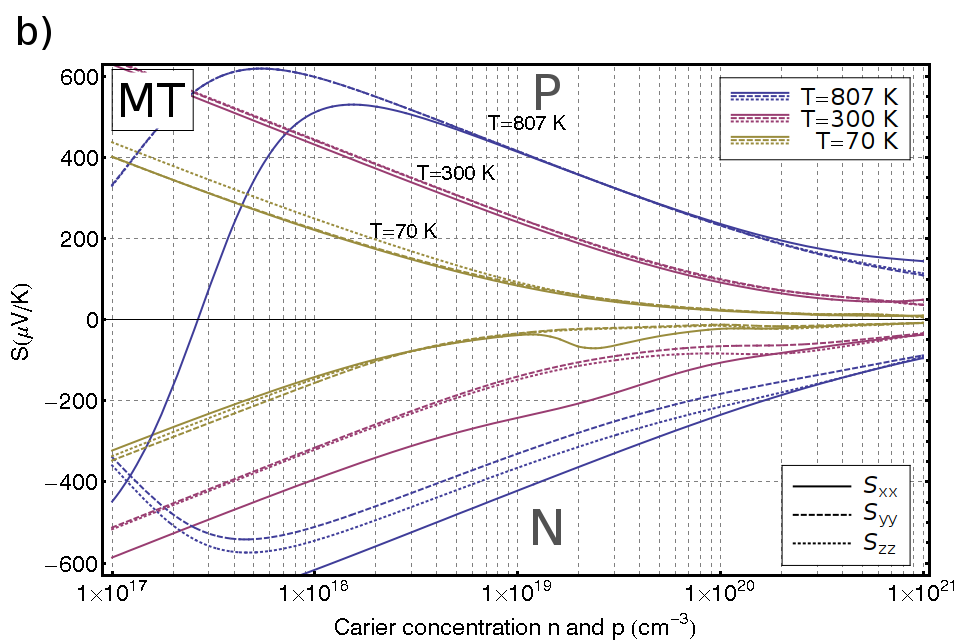}
\includegraphics[width=0.499\textwidth]{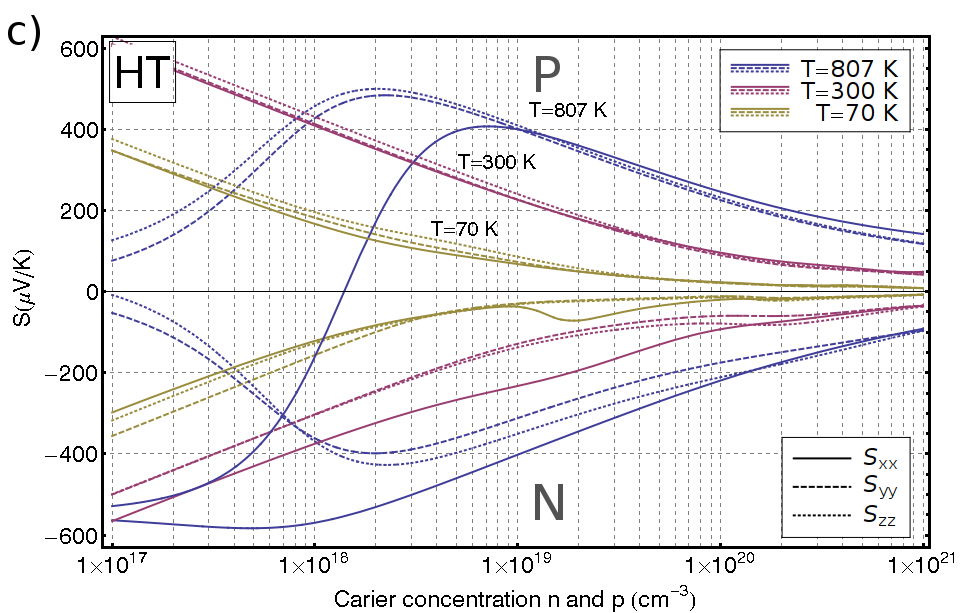}
\caption{(Color online) Thermopower LT (a), MT (b) and HT (c) phases at three different temperature and crystallographic direction.}
\label{fig:S_n_extra}
\end{figure}

\begin{figure}[H]
\centering
\includegraphics[width=0.39\textwidth]{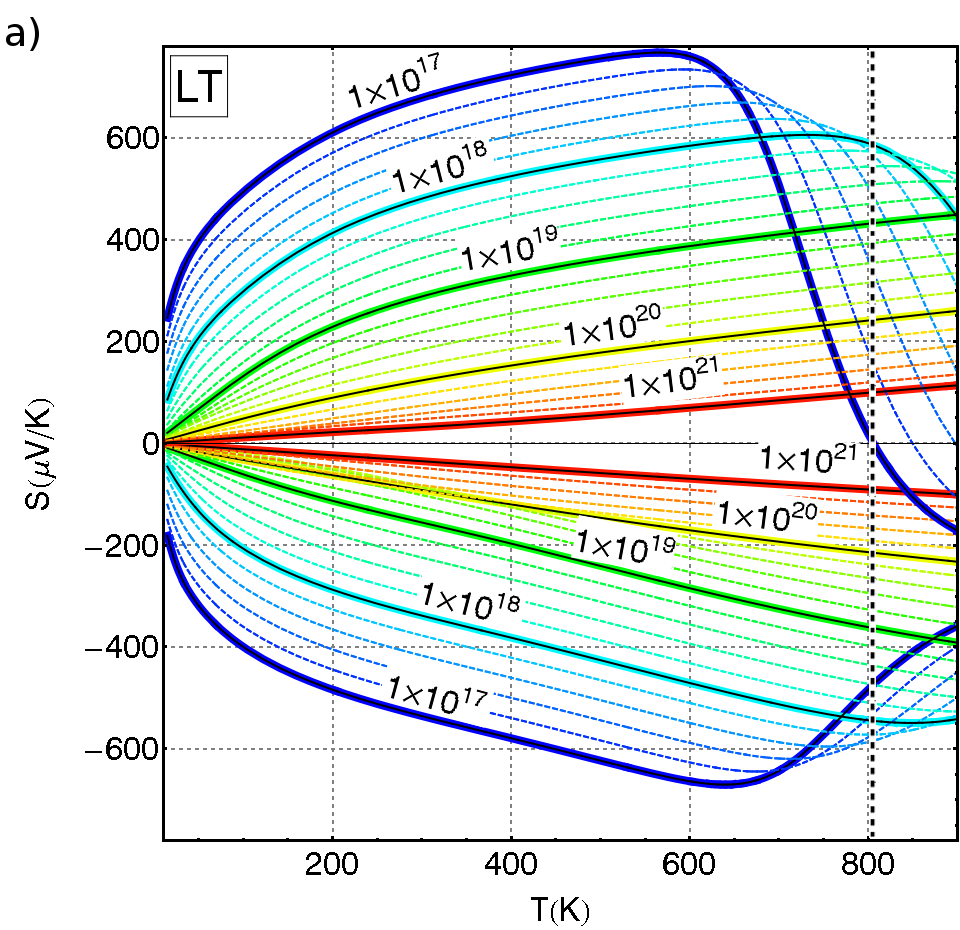}
\includegraphics[width=0.08\textwidth]{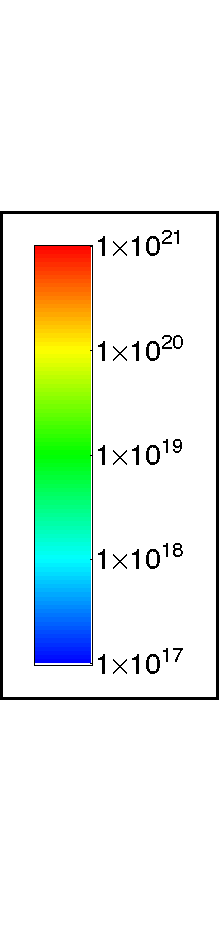}
\includegraphics[width=0.39\textwidth]{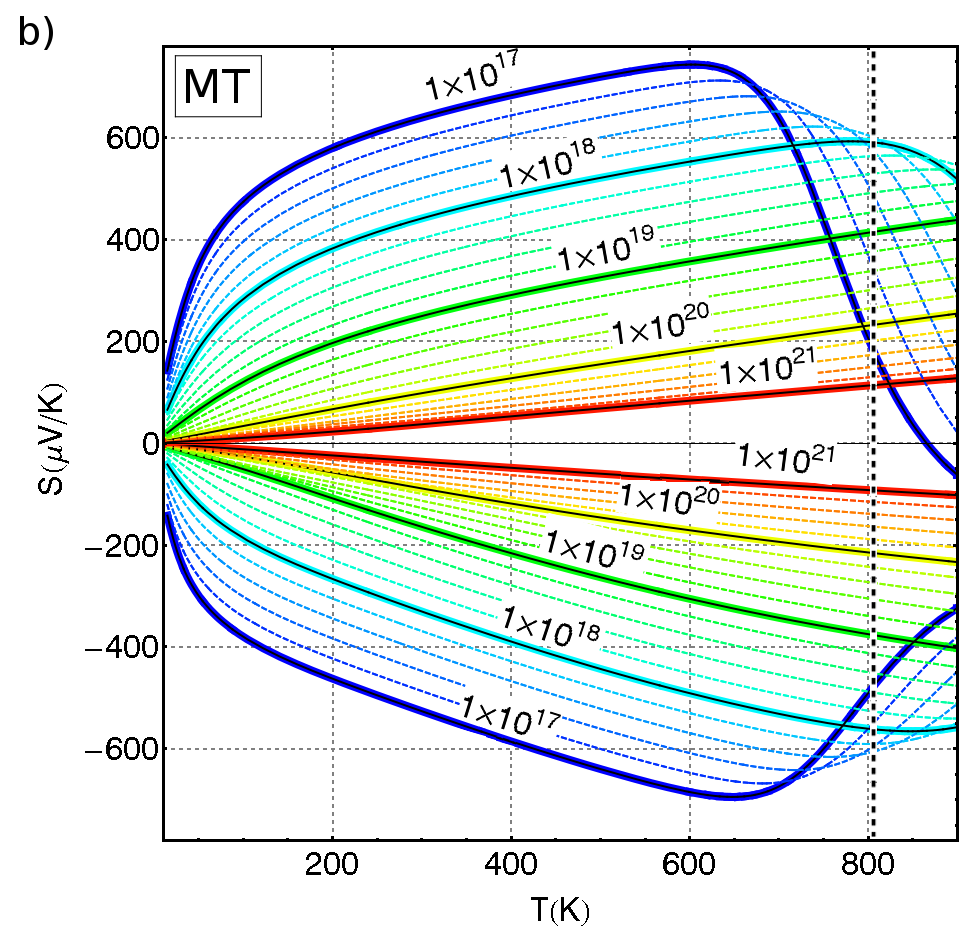}
\includegraphics[width=0.08\textwidth]{FIGS/legenda_v4.png}
\includegraphics[width=0.39\textwidth]{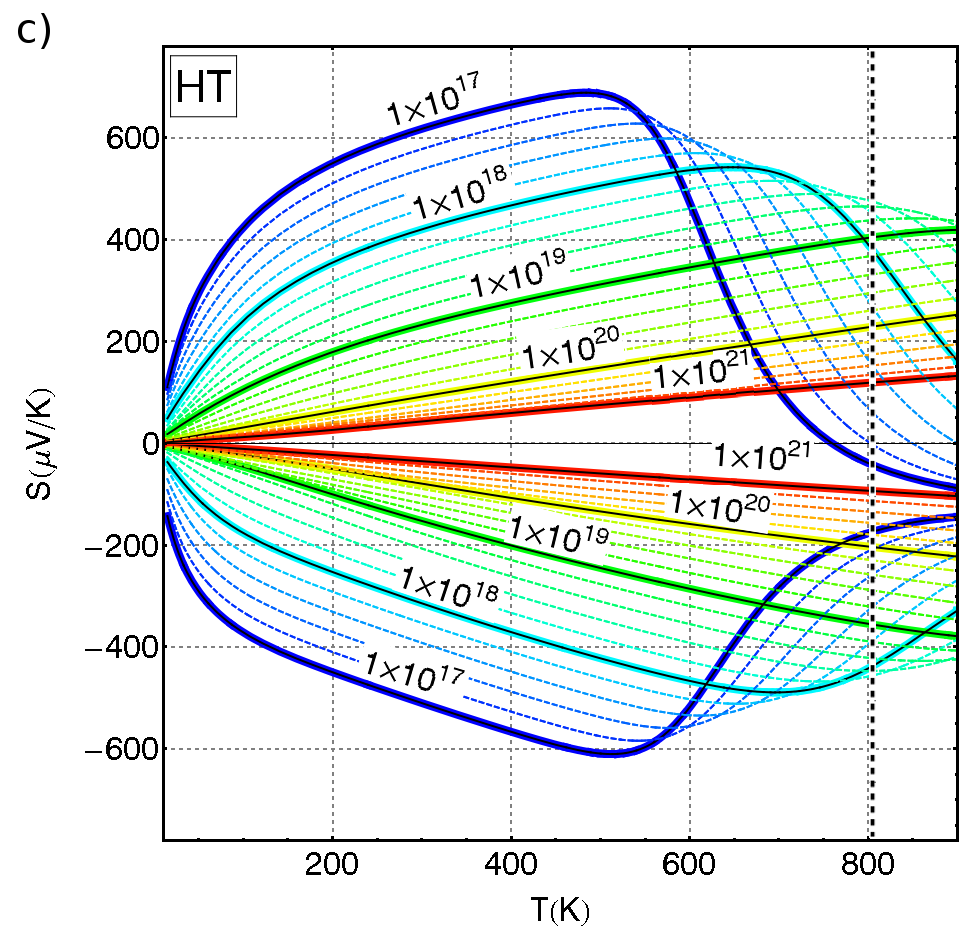}
\includegraphics[width=0.08\textwidth]{FIGS/legenda_v4.png}
\caption{(Color online)  Isotropic thermopower in function of temperature for LT (a), MT (b) and HT (c) for  the  $n$ and $p$-type doping. Different color lines correspond to  the  different carrier concentrations. Thick line marks concentrations in cm$^{-3}$.}
\label{fig:ST}
\end{figure}

In this Appendix, additional figures are included, showing the complete data for the anisotropic (Fig.~\ref{fig:S_n_extra}) and isotropic (Fig.~\ref{fig:ST}) thermopowers of the three considered SnSe structures. 

\begin{acknowledgments}
This work was supported by the Polish National Science Center (NCN) under the grants DEC-2011/02/A/ST3/00124 and partially by the Polish Ministry of Science and Higher Education.
\end{acknowledgments}



\end{document}